\documentclass[preprint,review,3p,times]{elsarticle}
\newcommand{\EQ}{\begin{equation}}
\newcommand{\EN}{\end{equation}}
\newcommand{\EQA}{\begin{eqnarray}}
\newcommand{\ENA}{\end{eqnarray}}
\usepackage[namelimits]{amsmath}
\usepackage{amssymb}
\usepackage{amsfonts}
\usepackage{mathrsfs}
\usepackage[colorlinks,linkcolor=blue]{hyperref}
\usepackage{graphicx}
\usepackage{makecell}
\biboptions{sort&compress}
\newcounter{bla}

\begin{document}

\begin{frontmatter}

%% Title, authors and addresses

%% use the tnoteref command within \title for footnotes;
%% use the tnotetext command for the associated footnote;
%% use the fnref command within \author or \address for footnotes;
%% use the fntext command for the associated footnote;
%% use the corref command within \author for corresponding author footnotes;
%% use the cortext command for the associated footnote;
%% use the ead command for the email address,
%% and the form \ead[url] for the home page:
%%
%% \title{Title\tnoteref{label1}}
%% \tnotetext[label1]{}
%% \author{Name\corref{cor1}\fnref{label2}}
%% \ead{email address}
%% \ead[url]{home page}
%% \fntext[label2]{}
%% \cortext[cor1]{}
%% \address{Address\fnref{label3}}
%% \fntext[label3]{}

\title{{\em DeepFlame}: A deep learning empowered open-source platform for reacting flow simulations}

%% use optional labels to link authors explicitly to addresses:
%% \author[label1,label2]{<author name>}
%% \address[label1]{<address>}
%% \address[label2]{<address>}

\author[a,b]{Runze Mao}
\author[a,b]{Minqi Lin}
\author[c,d]{Yan Zhang\corref{author}}
\author[e,b]{Tianhan Zhang}
\author[f,g,b]{Zhi-Qin John Xu}
\author[a,b]{Zhi X. Chen\corref{author}}

\cortext[author] {Corresponding author.\\\textit{E-mail addresses: } zhang\_yan2@iapcm.ac.cn (Y. Zhang), chenzhi@pku.edu.cn (Z.X. Chen)}
\address[a]{State Key Laboratory of Turbulence and Complex Systems, Aeronautics and Astronautics, College of Engineering, Peking University, Beijing, 100871, China}
\address[b]{AI for Science Institute, Beijing, 100080, China}
\address[c]{CAEP Software Center for High Performance Numerical Simulation, Beijing 100088, China}
\address[d]{Institute of Applied Physics and Computational Mathematics, Beijing 100088, China}
\address[e]{Department of Mechanics and Aerospace Engineering, SUSTech, Shenzhen, 518055, China}
\address[f]{Institute of Natural Sciences, School of Mathematical Sciences, Shanghai Jiao Tong University, Shanghai, 200240, China}
\address[g]{MOE-LSC and Qing Yuan Research Institute, Shanghai Jiao Tong University, Shanghai, 200240, China}

\begin{abstract}

%% Text of abstract
Recent developments in deep learning have brought many inspirations for the scientific computing community and it is perceived as a promising method in accelerating the computationally demanding reacting flow simulations. In this work, we introduce {\em DeepFlame}, an open-source C++ platform with the capabilities of utilising machine learning algorithms and pre-trained models to solve for reactive flows. We combine the individual strengths of the computational fluid dynamics library OpenFOAM, machine learning framework Torch, and chemical kinetics program Cantera. The complexity of cross-library function and data interfacing (the core of {\em DeepFlame}) is minimised to achieve a simple and clear workflow for code maintenance, extension and upgrading.
As a demonstration, we apply our recent work on deep learning for predicting chemical kinetics (Zhang et al. \textit{Combust. Flame vol.~245 pp.~112319, 2022}) to highlight the potential of machine learning in accelerating reacting flow simulation.
A thorough code validation is conducted via a broad range of canonical cases to assess its accuracy and efficiency. The results demonstrate that the convection-diffusion-reaction algorithms implemented in {\em DeepFlame} are robust and accurate for both steady-state and transient processes. In addition, a number of methods aiming to further improve the computational efficiency, e.g. dynamic load balancing and adaptive mesh refinement, are explored. Their performances are also evaluated and reported. With the deep learning method implemented in this work, a speed-up of two orders of magnitude is achieved in a simple hydrogen ignition case when performed on a medium-end graphics processing unit (GPU). Further gain in computational efficiency is expected for hydrocarbon and other complex fuels. A similar level of acceleration is obtained on an AI-specific chip – deep computing unit (DCU), highlighting the potential of {\em DeepFlame} in leveraging the next-generation computing architecture and hardware.

\end{abstract}


\end{frontmatter}

%%
%% Start line numbering here if you want
%%
% \linenumbers

% Computer program descriptions should contain the following
% PROGRAM SUMMARY.

{\bf PROGRAM SUMMARY}
  %Delete as appropriate.

\begin{small}
\noindent
{\em Program Title:} DeepFlame                                          \\
{\em CPC Library link to program files:} (to be added by Technical Editor)   \\
  %Leave blank, supplied by Elsevier.
{\em Developer's repository link:} \url{https://github.com/deepmodeling/deepflame-dev} \\
{\em code Ocean capsule:} (to be added by Technical Editor)   \\
  %Leave blank, supplied by Elsevier.
{\em Licensing provisions:} GPLv3                                  \\
  %enter "none" if CPC non-profit use license is sufficient.
{\em Programming language:} C++                               \\
% {\em Computer:}                                               \\
  %Computer(s) for which program has been designed.
% {\em Operating system:}                                       \\
  %Operating system(s) for which program has been designed.
%{\em RAM:} bytes                                              \\
  %RAM in bytes required to execute program with typical data.
%{\em Number of processors used:}                              \\
  %If more than one processor.
%{\em Supplementary material:}                                 \\
  % Fill in if necessary, otherwise leave out.
%{\em Keywords:} Keyword one, Keyword two, Keyword three, etc.  \\
  % Please give some freely chosen keywords that we can use in a
  % cumulative keyword index.
%{\em Classification:}                                         \\
  %Classify using CPC Program Library Subject Index, see (
  % http://cpc.cs.qub.ac.uk/subjectIndex/SUBJECT_index.html)
  %e.g. 4.4 Feynman diagrams, 5 Computer Algebra.
%{\em External routines/libraries:}                            \\
  % Fill in if necessary, otherwise leave out.
%{\em Subprograms used:}                                       \\
  %Fill in if necessary, otherwise leave out.
%{\em Catalogue identifier of previous version:}*              \\
  %Only required for a New Version summary, otherwise leave out.
%{\em Journal reference of previous version:}*                  \\
  %Only required for a New Version summary, otherwise leave out.
%{\em Does the new version supersede the previous version?:}*   \\
  %Only required for a New Version summary, otherwise leave out.
{\em Nature of problem:} Solving chemically reacting flows with direct (quasi-direct) simulation methods is usually troubled by the following problems: 1. as the widely-used CFD toolbox, OpenFOAM features poor ODE solvers for chemistry and oversimplified transport models, yielding non-negligible errors in simulation results; 2. the chemical source term evaluation is the most computationally expensive and usually accounts for more than 80$\%$ of total computing time.\\
  %Describe the nature of the problem here.
{\em Solution method:} An open-source platform bringing together the individual strengths of OpenFOAM, Cantera and PyTorch libraries is built in this study. In the present implementation, CVODE solvers, detailed transport models and deep learning algorithms are all adopted to assist the simulation of reacting flow. Note that here machine learning is introduced in combination with heterogeneous computing to accelerate the most demanding solving procedure for chemical source term evaluation.\\
\end{small}
  %Describe the method solution here.
%   \\
%{\em Reasons for the new version:}*\\
  %Only required for a New Version summary, otherwise leave out.
%   \\
%{\em Summary of revisions:}*\\
  %Only required for a New Version summary, otherwise leave out.
%   \\
%{\em Restrictions:}\\
%   Describe any restrictions on the complexity of the problem here.
%   \\
% {\em Unusual features:}\\
  %Describe any unusual features of the program/problem here.
%   \\
% {\em Additional comments:}\\
  %Provide any additional comments here.
%   \\
% {\em Running time:}\\
  %Give an indication of the typical running time here.
%   \\
%% main text
\section{Introduction}\label{sec:introd}

Computational Fluid Dynamics (CFD) tools for simulating reacting flows are crucial in developing less-polluting and highly-efficient energy and propulsion technologies \cite{poinsot2005theoretical}. While the turbulent flow and chemical reactions are strongly coupled in practical devices, modelling the multi-scale, multi-phase and multi-species physio-chemical processes under engine-relevant conditions remain a scientific challenge.  In addition, simulating reacting flows in strong turbulence with the scales and species fully resolved (known as direct numerical simulation, DNS) requires quite demanding computational resources. Simplified modelling methods such as Reynolds-averaged Navier-Stokes (RANS) approach and large eddy simulation (LES) mostly rely on statistical or topological models to impose physical assumptions to accelerate simulations. However, the generalisation abilities of these models have long been the major issue limiting the practical application of reacting flow simulations \cite{peters2000turbulent}. 

To resolve the above dilemma of accuracy versus efficiency, the recent rapid growth in Artificial Intelligence (AI) for Science, particularly in machine learning has brought new perspectives for accelerating simulation of reactive flows with accurate models and detailed chemistry. As a pioneer work, Christo et al. \cite{christo1996artificial} adopted Artificial neural network (ANN) in the joint PDF/Monte Carlo simulation of H$_2$/CO$_2$ turbulent jet diffusion flames to predict chemical kinetics. Blasco et al. \cite{blasco1998modelling} trained more than one ANNs based on a typical combustion simulation to capture the changes of species composition at various time steps, so that the reaction rates can be directly obtained by ANNs instead of solving ordinary differential equations (ODEs) or accessing look-up table. Sen et al. \cite{sen2009turbulent} successfully adopted ANNs as a chemical kinetic integrator for LES of turbulent flame. They found ANN exhibits satisfying behaviour both in memory and time efficiency. Wan et al. \cite{wan2020chemistry} trained a deep neural network (DNN) based on the turbulent micro-mixing data to predict the reaction rates. The DNN was used in simulating a turbulent non-premixed syngas oxy-flame and obtained a considerable speed-up. Yao et al. \cite{yao2022gradient} adopted gradient boosted decision tree (GBDT) as a machine learning approach to directly solve the chemistry ODEs and gained a speed-up of one order of magnitude. 
More recently, to improve the generalisation ability of DNN, Zhang et al. \cite{zhang2022multi} proposed a new sampling method for collecting multi-scale combustion data. The neural network trained from such data set has been confirmed to be accurate and efficient in predicting reaction rates under various conditions. Besides, DNN was also extended to the high-dimensional tabulation of flamelets and effectively reduced the memory requirement \cite{chen2021application,chi2022efficient,perry2022co,zhang2020large}. 

It is widely acknowledged that the growing success of deep learning is built upon open-source and data sharing culture in the scientific and industrial communities.
However, to the best of our knowledge, most of the studies mentioned above were conducted using in-house codes~\footnote{In some papers the training and testing codes were released in supplementary material or uploaded onto GitHub for reproducibility. However, the CFD codes in which the pre-trained model deployment and the following \textit{a posteriori} assessment were carried out remain difficult to access.}. Despite the promising potential of machine learning in this field, the related works are still limited because of the lack of the powerful tools and platforms leveraging the growing assets in fluid dynamics, machine learning and chemical kinetics communities. Nowadays, the advantages of open-source community in developing new technology has been widely proved. A number of open-source frameworks for machine learning (e.g. TensorFlow\cite{abadi2016tensorflow} and Torch\cite{collobert2011torch7}), CFD (e.g. OpenFOAM\cite{opencfd2009open}), and chemical kinetics computation (e.g. Cantera\cite{goodwin2002cantera}) are available for users and developers from both the scientific and industrial communities. However, it is necessary to build a reacting flow simulation platform that brings together the individual strengths of the CFD, machine learning and chemical kinetics open-source communities, so that the cross-disciplinary research interaction and code development can be facilitated.

With this motivation, the main objective of the present work is to develop an open-source CFD platform named {\em DeepFlame} for simulating reacting flows with capabilities of utilising state-of-the-art machine learning algorithms and libraries. Briefly, {\em DeepFlame} integrates existing libraries and organises the computational tasks in the following manner. Tools and functions for solving general continuum fluid flow problems are called or derived from OpenFOAM, including flow field data-structure, numerical discretisation, iteration for linear solvers, MPI-based parallel computing, and pre-/post-processing; chemical mechanism related I/O and multi-species thermochemistry  data-structure and property calculation are handled by Cantera; the deep learning framework libTorch (Torch C++ API) is coupled in {\em DeepFlame} for the manipulation of input/output tensor-format data and inference of deep neural network models. 
In addition, methods aiming to improve simulation efficiency such as dynamic load balance (DLB) and adaptive mesh refinement (AMR) have also been implemented. A preliminary heterogeneous computing approach is available in this version, where AI acceleration infrastructure (i.e. GPU) can be used to further magnify the simulation speed-up when machine learning models are activated. 

The structure of this paper is as follows. In Section \ref{sec:Govern}, we discuss the governing equations solved by the different flow solvers implemented in {\em DeepFlame}, where the approaches for obtaining chemistry source terms are emphasised. In Section \ref{sec:Inplem}, we introduce the implementation details including the code structure and algorithms. In Section \ref{sec:Res}, we conduct a broad range of canonical cases to validate the solvers for different flow conditions. In Section \ref{sec:Perf}, the computational performance of {\em DeepFlame} is evaluated. Finally, the conclusions and further works are summarised in Section \ref{sec:Conclusion}.

\section{Theoretical Background}\label{sec:Govern}
\subsection{Governing Equations}\label{subs:GorvEq}
To directly solve the compressible reacting flows with $N$ number of species, the conservation equations of mass, momentum, species and energy used in this work, in common tensorial notations, are given by

\EQ
\frac{\partial \rho}{\partial t} + \frac{\partial \rho u_i}{\partial x_i} =0 \:,
\EN
\EQ
\frac{\partial \rho u_j}{\partial t} + \frac{\partial \rho u_iu_j}{\partial x_i}=-\frac{\partial p}{\partial x_j}+ \frac{\partial \tau  _{ij}}{\partial x_i}  \:,
\label{eq:momenteq}
\EN
\EQ
\frac{\partial \rho Y_\alpha}{\partial t} +\frac{\partial \rho u_i Y_\alpha}{\partial x_i} =- \frac{\partial \rho Y_\alpha V_{\alpha,i}}{\partial x_i} + \dot{\omega}_\alpha  \:,
\label{eq:specieeq}
\EN
\begin{subequations}
\begin{align}
    \frac{\partial (\rho H)}{\partial t} +\frac{\partial (\rho u_i H)}{\partial x_i} & =\frac{\partial p}{\partial t}-\frac{\partial q_i}{\partial x_i} + \frac{\partial}{\partial x_j}(\tau _{ij} u_i) \:,\\
     {\rm and}~~~~\frac{\partial (\rho E)}{\partial t} +\frac{\partial (\rho u_i E)}{\partial x_i} & =-\frac{\partial q_i}{\partial x_i} + \frac{\partial}{\partial x_j}\left[(\tau _{ij}-p\delta_{ij}) u_i\right]  \:,
\end{align}
\label{eq:energyeq}
\end{subequations}
where $t$ is time, $u_j$ and $x_j$ are the velocity component and Cartesian spatial coordinate in the $j$ direction respectively, $\rho$ is mixture mass density, and $p$ is pressure. In Eq.~\eqref{eq:specieeq}, $Y_\alpha$ and $\dot{\omega}_\alpha$ are mass fraction and net reaction rate of the $\alpha$-th species, respectively. In Eq.~\eqref{eq:energyeq}, $H = h + \frac{1}{2}u_i u_i$ and $E = e + \frac{1}{2}u_i u_i$ are the total enthalpy (absolute enthalpy + kinetic energy) and total energy (internal energy + kinetic energy) respectively, and $q_i$ is energy flux in direction $i$.
Depending on the flow velocity with respect to the speed of sound, different forms of energy are solved and this will be described in detail later in §\ref{subs:Solvers}.

The viscous tensor $\tau_{ij}$ in Eq.~\eqref{eq:momenteq}, by applying Stokes' hypothesis, is written as
\EQ
\tau _{ij}=-\frac{2}{3}\mu \frac{\partial u_k}{\partial x_k}  \delta _{ij}+\mu \left(\frac{\partial u_i}{\partial x_j}+\frac{\partial u_j}{\partial x_i}\right)\:.
\EN
The dynamic viscosity of mixture, $\mu$, is calculated via the Wilke mixture formulation:
\EQ
\mu = \sum_{\alpha}\frac{\mu _\alpha X_\alpha}{ {\textstyle \sum_{\beta}}\Phi _{\alpha,\beta}X_\beta } 
\EN
with $\mu_\alpha$ and $X_\alpha$ being the dynamic viscosity and mole fraction of the $\alpha$-th species respectively, and $\Phi_{\alpha,\beta}$ is computed using
\EQ
\Phi _{\alpha,\beta}=\frac{\left [ 1+\sqrt{\left ( \frac{\mu _\alpha}{\mu _\beta}\sqrt{\frac{M_\beta}{M_\alpha} }   \right ) }  \right ]^2 }{\sqrt{8}\sqrt{1+M_\alpha/M_\beta}} \:,
\label{eq:Phi}
\EN
where $M_\alpha$ is the mole mass of the $\alpha$-th species.

In the species transport equations (see Eq.~(\ref{eq:specieeq})), the reaction rates, $\dot{\omega}_\alpha$, can be obtained from the Arrhenius Law and on-the-fly integration or a pre-trained deep neural network, which will be introduced in §~\ref{subs:ChemEq}. 
The diffusion velocity of the $\alpha$-th species in $i$-th direction, $V_{\alpha,i}$, is given by Fick's Law and corrected with a correction velocity $V_i^c$ to ensure mass conservation:
\EQ
V_{\alpha,i}=-\frac{D_\alpha}{Y_\alpha}\frac{\partial Y_\alpha}{\partial x_i}+\underset{V_i^c}{ \underbrace{\sum_{\alpha=1}^{N} (D_\alpha\frac{\partial Y_\alpha}{\partial x_i}) } } \:,
\label{eq:V_ki}
\EN
where $D_\alpha$ is the diffusion coefficient between $\alpha$-th species and the rest of the mixture. By default, {\em DeepFlame} employs (via a Cantera interface function) the Hirschfelder and Curtiss mixture-averaged transport model \cite{curtiss1949transport}:
\EQ
D_\alpha = \frac{1-Y_\alpha}{ {\textstyle \sum_{\beta\ne \alpha}}X_\beta/\mathcal{D}_{\beta\alpha}} \:,
\EN
where $\mathcal{D}_{\beta\alpha}$ is the binary diffusion coefficient between $\alpha$-th species and $\beta$-th species, which can be obtained according to Takahashi correlation \cite{kee2005chemically}. The simpler model implemented in OpenFOAM assuming unity Lewis number ($Le$) for all species is also retained in {\em DeepFlame}, and the mass diffusivity is modelled as 
\EQ
D_\alpha \equiv \frac{a}{Le} = \frac{\lambda}{\rho C_p}\:,
\label{eq:D_UL}
\EN
where $a$, $\lambda$ and $C_p$ are the thermal diffusivity, thermal conductivity and constant-pressure heat capacity of the mixture, respectively.
The mixture overall value of $\lambda$ is obtained using the conductivity of each pure species ($\lambda_\alpha$):
\EQ
\lambda =0.5(\sum_{\alpha}X_\alpha\lambda_\alpha+\frac{1}{ {\textstyle \sum_{\alpha}X_\alpha/ \lambda_\alpha} } ) \:,
\label{eq:lambda}
\EN
and a similar procedure is used for $C_p$ based on the JANAF database.
The more detailed multi-component transport model is also supported via the Cantera interface function and the theoretical details are available in \cite{ern1994lecture}.

The energy equation for total enthalpy and total energy are respectively given by Eq.~(\ref{eq:energyeq}a) and (\ref{eq:energyeq}b), the energy flux $q_i$ is
\EQ
q_i = -\lambda \frac{\partial T}{\partial x_i} + \rho \sum_{\alpha=1}^{N} h_\alpha Y_\alpha V_{\alpha,i}
\label{eq:qi} \:.
\EN
The RHS of the above equation includes a heat conduction term (expressed by Fourier's Law, $\lambda {\partial T}/{\partial x_i}$) and a species diffusion term, where $h_\alpha$ is the enthalpy of the $\alpha$-th species.

Furthermore, temperature in Eq.~(\ref{eq:qi}) is usually cast to enthalpy to facilitate the numerical solving procedure of the energy equation. The heat conduction term is rewritten as
\EQ
\begin{aligned}
\lambda \frac{\partial T}{\partial x_i} &= \rho a \sum_{\alpha=1}^{N}Y_\alpha C_{p\alpha}\frac{\partial T}{\partial x_i} \\
&= \sum_{k=1}^{N}\rho a Y_\alpha\frac{\partial h_\alpha}{\partial x_i}\\
&= \rho a \frac{\partial h}{\partial x_i} - \sum_{\alpha=1}^{N}\rho a h_\alpha\frac{\partial Y_\alpha}{\partial x_i}
\end{aligned}
\label{eq:gradT}
\EN

Finally, combined with Eq.~(\ref{eq:V_ki}), Eq.~(\ref{eq:qi}) and  Eq.~(\ref{eq:gradT}), the energy equation (Eq.~(\ref{eq:energyeq}a) and (\ref{eq:energyeq}b)) can be expressed more detailed as:
\begin{subequations}
\begin{align}
&\frac{\partial (\rho H)}{\partial t} +\frac{\partial (\rho u_i H)}{\partial x_i}=\frac{\partial p}{\partial t}+ \frac{\partial}{\partial x_j}(\tau _{ij} u_i)+\frac{\partial }{\partial x_i} (\rho a \frac{\partial h}{\partial x_i} ) \nonumber \\ &-\sum_{\alpha=1}^{N}\frac{\partial}{\partial x_i} (\rho a h_\alpha\frac{\partial Y_\alpha}{\partial x_i} )-\frac{\partial}{\partial x_i}\left [ \rho\sum_{\alpha=1}^{N}h_\alpha Y_\alpha(-\frac{D_\alpha}{Y_\alpha}\frac{\partial Y_\alpha}{\partial x_i} +\sum_{\alpha=1}^{N}D_\alpha\frac{\partial Y_\alpha}{\partial x_i}) \right ] \\
&\frac{\partial (\rho E)}{\partial t} +\frac{\partial (\rho u_i E)}{\partial x_i}=\frac{\partial}{\partial x_i}\left [(\tau _{ij}-p\delta_{ij}) u_i\right ]+\frac{\partial }{\partial x_i} (\rho a \frac{\partial h}{\partial x_i} ) \nonumber \\ &-\sum_{\alpha=1}^{N}\frac{\partial}{\partial x_i} (\rho a h_\alpha\frac{\partial Y_\alpha}{\partial x_i} )-\frac{\partial}{\partial x_i}\left [ \rho\sum_{\alpha=1}^{N}h_\alpha Y_\alpha(-\frac{D_\alpha}{Y_\alpha}\frac{\partial Y_\alpha}{\partial x_i} +\sum_{\alpha=1}^{N}D_\alpha\frac{\partial Y_\alpha}{\partial x_i}) \right ]
\end{align}
\label{eq:energy2}
\end{subequations}

\subsection{DeepFlame Solvers}\label{subs:Solvers}
Based the above governing equations, {\em DeepFlame} provides three solvers to simulate reacting flow under different conditions:
\begin{itemize}
    \item {\em df0DFoam}-solver: developed to solve zero-dimensional auto-ignition problems, concerning both constant-pressure and constant-volume cases. The convection and diffusion terms are discarded here, and thus only the species equations are solved to estimate the time-variation of the thermochemical states of the zero-dimensional reactor. 
    \item {\em dfLowMachFoam}-solver: developed based on {\em rhoPimpleFoam} (the original pressure-based compressible solver in OpenFOAM) for simulation of low-Mach number reacting flows. The equation for total enthalpy (Eq.~(\ref{eq:energy2}a)) is solved here to describe conservation of energy, in which the contribution of viscous heating (the second term on the RHS) is neglected. Additionally, the Strang splitting scheme \cite{strang1968construction} is adopted to improve the solving accuracy.
    \item {\em dfHighSpeedFoam}-solver: developed based on {\em rhoCentralFoam} (the original density-based compressible solver in OpenFOAM) solving algorithm to simulate high-speed reacting flows. The equation for total energy (Eq.~(\ref{eq:energy2}b)) is applied to describe the conservation of energy. The viscous heating and correction of diffusion velocity are taken into account when the flow is considered to be viscous
\end{itemize}   

\subsection{Chemistry Integration}\label{subs:ChemEq}
Typically, in the simulation of reacting flows, the computational cost of the chemistry source term evaluations is completely dominant and even exceeds the  cost of fluid dynamics by a factor of 100 \cite{peters2000turbulent}. Therefore, improving the efficiency of chemistry solver is crucial in the development of advanced reacting flow simulation platforms \cite{lu2009toward}. In this section, the methods evaluating the chemistry source term in {\em DeepFlame} are introduced. Here we start with a chemical system of $N$ species and $M$ reactions \cite{poinsot2005theoretical}:
\EQ
\sum_{\alpha=1}^{N} {\nu}'_{\alpha j} \mathcal{M}_\alpha\rightleftharpoons \sum_{\alpha=1}^{N} {\nu}''_{\alpha j} \mathcal{M}_\alpha \ \ \ \mathrm{for} \ \ \ j=1,\dots,M\:,
\EN
where $\mathcal{M}_\alpha$ denotes species $\alpha$, ${\nu}'_{\alpha j}$ and ${\nu}''_{\alpha j}$ represent the molar stoichiometric coefficients of species $\alpha$ in reaction $j$. The reaction rate of species $\alpha$ in such a system is calculated from
\EQ
\frac{\mathrm{d} Y_\alpha}{\mathrm{d} t} = W_\alpha\sum_{j=1}^{M} (\nu''_{\alpha j}-\nu'_{\alpha j})\left \{ K_{fj} {\textstyle \prod_{\alpha=1}^{N}}(\frac{\rho Y_\alpha}{W_\alpha})^{\nu '_{\alpha j}} -K_{rj} {\textstyle \prod_{\alpha=1}^{N}} (\frac{\rho Y_\alpha}{W_\alpha})^{\nu ''_{\alpha j}} \right \} \:,
\label{eq:sourceeq}
\EN
where $K_{fj}$ and $K_{rj}$ are the forward and reverse rates of reaction $j$, which are usually modelled using the well-known Arrhenius law. The molecular weight and molar concentration are denoted by $W_\alpha$ and $X_\alpha$ for the $\alpha$-th species, respectively. 

Typically, the chemical source term $\dot{\omega}_\alpha$ in Eq.~(\ref{eq:specieeq}) is solved based on the large ODEs system formed by Eq.~(\ref{eq:sourceeq}). It has been shown in \cite{RN12,RN5,RN13,RN14} that the default OpenFOAM ODE solver suffers from several drawbacks (e.g. poor accuracy and stability, slow integration, etc.). Following these previous studies , {\em DeepFlame} provides an interface function to call the well-established implicit ODE solver (CVODE of the SUNDIALS package \cite{hindmarsh2005sundials} from Cantera) to improve the solving accuracy and efficiency for the computation of reaction rates. 

\begin{figure}[!h]
\centering
\includegraphics[scale=0.6]{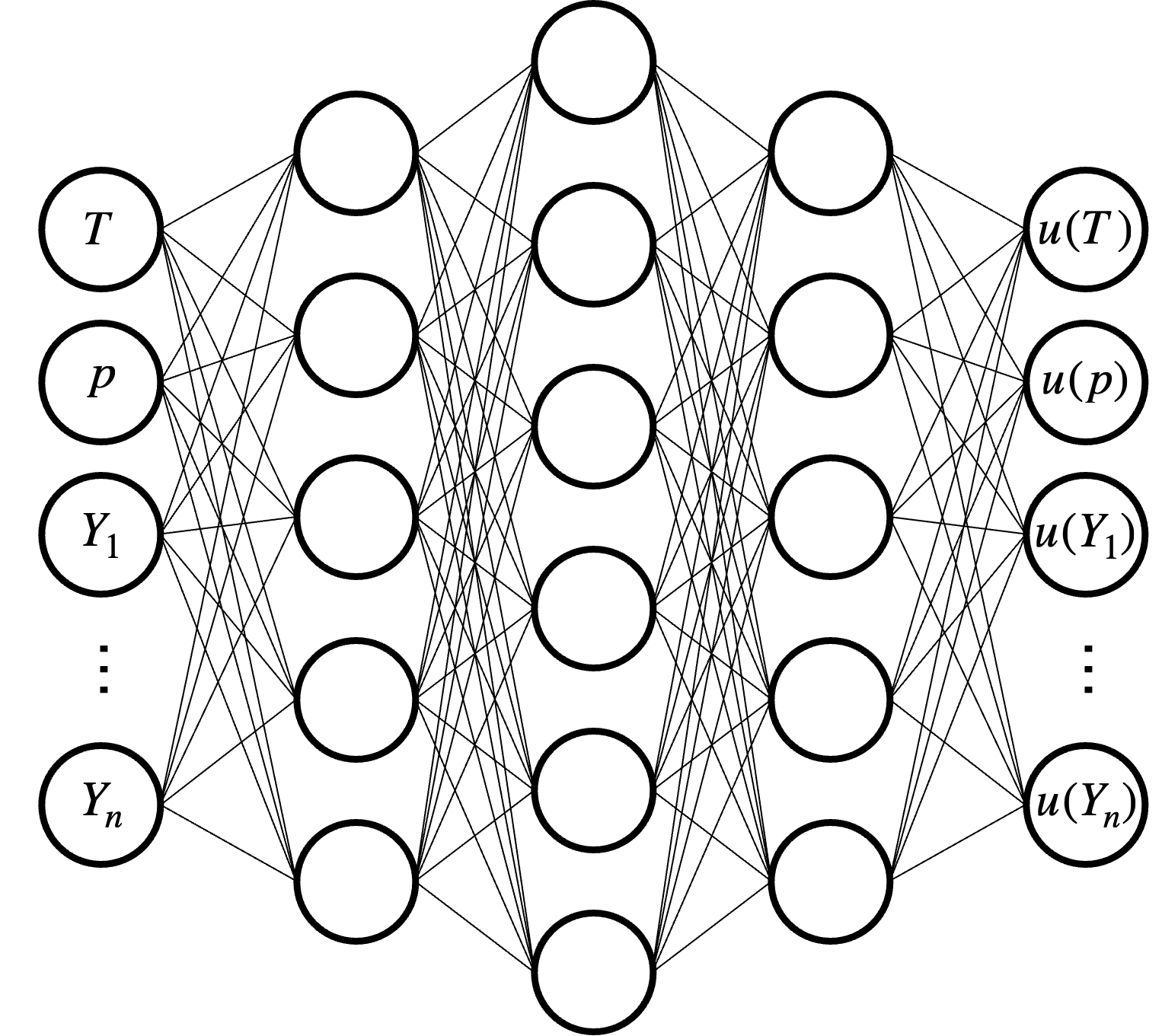}
\caption{The schematics of the DNN configurations.}
\label{fig:DNN}
\end{figure} 

However, since the time scale in reaction system has a broad distribution (from $\mathcal{O}$(ns) to $\mathcal{O}$(s)), the ODEs system is so stiff that the computational cost of which is still dominant in simulating reacting flow. Therefore, {\em DeepFlame} also adopts machine learning method to accelerate the solution of chemistry. Following the approach proposed by Zhang et al. \cite{zhang2022multi}, a fully-connected deep neutral network with three hidden layers is trained. The architecture of the DNN model is schematically demonstrated in Fig.~\ref{fig:DNN}. The input vector $\boldsymbol{x}(t)=\{T(t),P(t),\mathcal{F}(Y_\alpha(t))_{\alpha=1, \dots, n}\}$ represents the temperature, pressure and mass fractions of each species at time $t$. The operator $\mathcal{F}()$ donates the Box-Cox transformation (BCT) \cite{box1964analysis} which aims to convert the mass fractions from low-order quantity to $\mathcal{O}$(1). The output of the DNN (labelled as $\boldsymbol{u}(\boldsymbol{x})$) is the change of the input $\boldsymbol{x}$ during a large time step (typically $\Delta t=1\ \mu s$). Now, the chemistry source term $\dot{\omega}_\alpha$ can be explicitly obtained via $\boldsymbol{u}(Y_\alpha)$ and thus DNN can be regarded as an ODE integrator. Later in Section \ref{sec:Res} and Section \ref{sec:Perf}, the accuracy and computational efficiency of this deep learning method will be demonstrated in various cases.

\section{Implementation Details}\label{sec:Inplem}
\subsection{DeepFlame Code Structure}

\begin{figure}[!h]
\centering
\includegraphics[width=\textwidth]{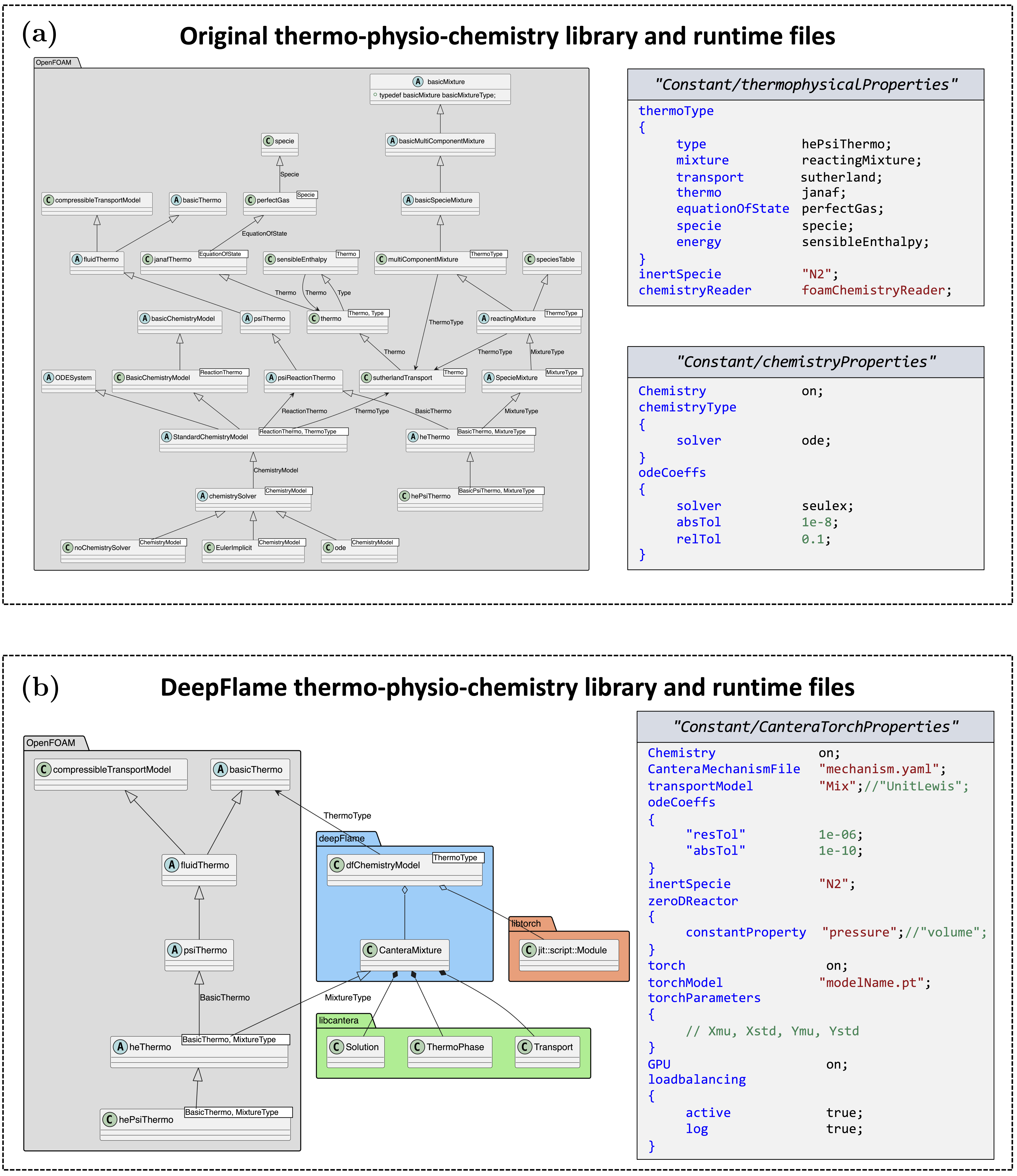}
\caption{The class diagrams of thermo-physio-chemistry library in (a) OpenFOAM and (b) {\em DeepFlame}. Lines ending with hollow triangle, hollow diamond and filled diamond shapes denote the inheritance, aggregation and composition relationships, respectively. Note that a line with sharp arrow denotes one class is adopted as a template in another. The corresponding case setup files are shown on the right.}
\label{fig:UML}
\end{figure} 
The first distinct feature of {\em DeepFlame} as compared to the existing codes having an OpenFOAM-Cantera interface is that our implementation for the two-way coupling is more compact and the cleanest possible. All the complex thermo-physcio-chemistry operations are handled using Cantera functions via an interface class. The lengthy and redundant OpenFOAM native species, mixture, reaction and derived thermophysical classes have been removed, and only the base class {\em fluidThermo} is kept to couple with the flow solver (same as in {\em rhoPimpleFoam} for non-reacting flow). Specifically, the coupling of OpenFOAM and libCantera (Cantera C++ API) is achieved as follows: OpenFOAM solves the basic conservation equations and outputs state parameters ($p$, $T$, $Y_\alpha$); libCantera is responsible for handling the chemical mechanism as well as the calculation of thermophysical coefficients ($\mu$, $\lambda$, $D_\alpha$) and reaction rates ($\dot{\omega}_\alpha$). In {\em DeepFlame}, the coupling routine is implemented in a new thermo-physio-chemistry library, which strictly follows the standard classes in OpenFOAM. In the following, a comparison between the newly developed and the original libraries is illustrated, and the advantageous features of {\em DeepFlame} can be easily observed. 

First, the classes related to the thermo-physio-chemistry library of OpenFOAM are summarized. The class diagram as well as  the corresponding runtime files (interface for user settings) are shown in Fig.~\ref{fig:UML}a. As listed in the class diagram, OpenFOAM develops a large number of classes with complicated relationship to describe the thermal and chemical properties of the mixture. The abstract basic classes are constructed to declare public interfaces. For example, the functions returning the field thermal properties are virtually defined in {\em basicThermo}, the mixture-related functions which output the thermal properties of mixture in a single cell are declared in {\em basicSpecieMixture}. These virtual functions are implemented in the derived classes with the models specified in runtime files by users. In general, the thermophysical models are determined in the file ``{\em constant/thermophysicalProperties}" and the chemistry models are set in the file ``{\em constant/chemistryProperties}". Here we produce a brief introduction of these models via a setting example illustrated in the right part of Fig.~\ref{fig:UML}a. In {\em thermoType}-dict, the keyword {\em type} specifies the underlying thermophysical model is {\em rho}-based or {\em psi}-based, reacting or nonreacting; the keyword {\em mixture} specifies the mixture composition is fixed or variable; keywords {\em transport} and {\em thermo} determine the transport and thermodynamic models, which are used in evaluating transport-parameters ($\mu$, $\lambda$ and $a$) and specific heat $C_p$; the last three keywords represent the equations of state model, species model (calculates the composition of each constituent) and energy form (internal energy or enthalpy), respectively. The {\em chemistryType}-dict specifies the solver for chemistry kinetics ODEs, and the coefficients of which is set in {\em odeCoeffs}-dict.

Figure~\ref{fig:UML}b illustrates the simplified structure of thermo-physio-chemistry library in {\em DeepFlame}. From the class diagram, an obvious simplification can be noted:
only several {\em thermo} classes derived from {\em basicThermo} are kept; the {\em mixture} classes and the {\em chemistry} classes are all replaced by {\em CanteraMixture} and {\em dfChemistryModel}. {\em CanteraMixture} is a class built to return the thermophysical properties calculated via libCantera, while {\em dfChemistryModel} is a chemistry model which enables CVODE and deep neural network for the solution of chemistry kinetics ODEs. Therefore, libCantera and libTorch are also included in this library. As shown in Fig.~\ref{fig:UML}b, the settings of the two classes are all contained in a new file ``{\em constant/CanteraTorchProperties}'': keyword {\em CanteraMechanismFile} is defined to read chemistry mechanisms; keyword {\em transportModel} sets the transport model, {\em UnityLewis}, {\em Mix}, and {\em Multi} are provided for users; the {\em zeroDReactor}-dict specifies the conditions for zero-dimensional reactor; the switch {\em torch} controls the use of deep neural network, and the network name as well as the normalisation parameters are respectively set in {\em torchModel} and {\em torchParameters}; the device for the DNN inference can also be specified by user via the switch {\em GPU}; the keywords in {\em loadbalancing}-dic specify the load balance switch for solving chemistry and log file output. Next, in §~\ref{subs:models}, the detailed structure and the underling algorithm of the two classes will be further introduced. 

\subsection{Description of the new classes}\label{subs:models}

\begin{figure}[!h]
\centering
\includegraphics[width=\textwidth]{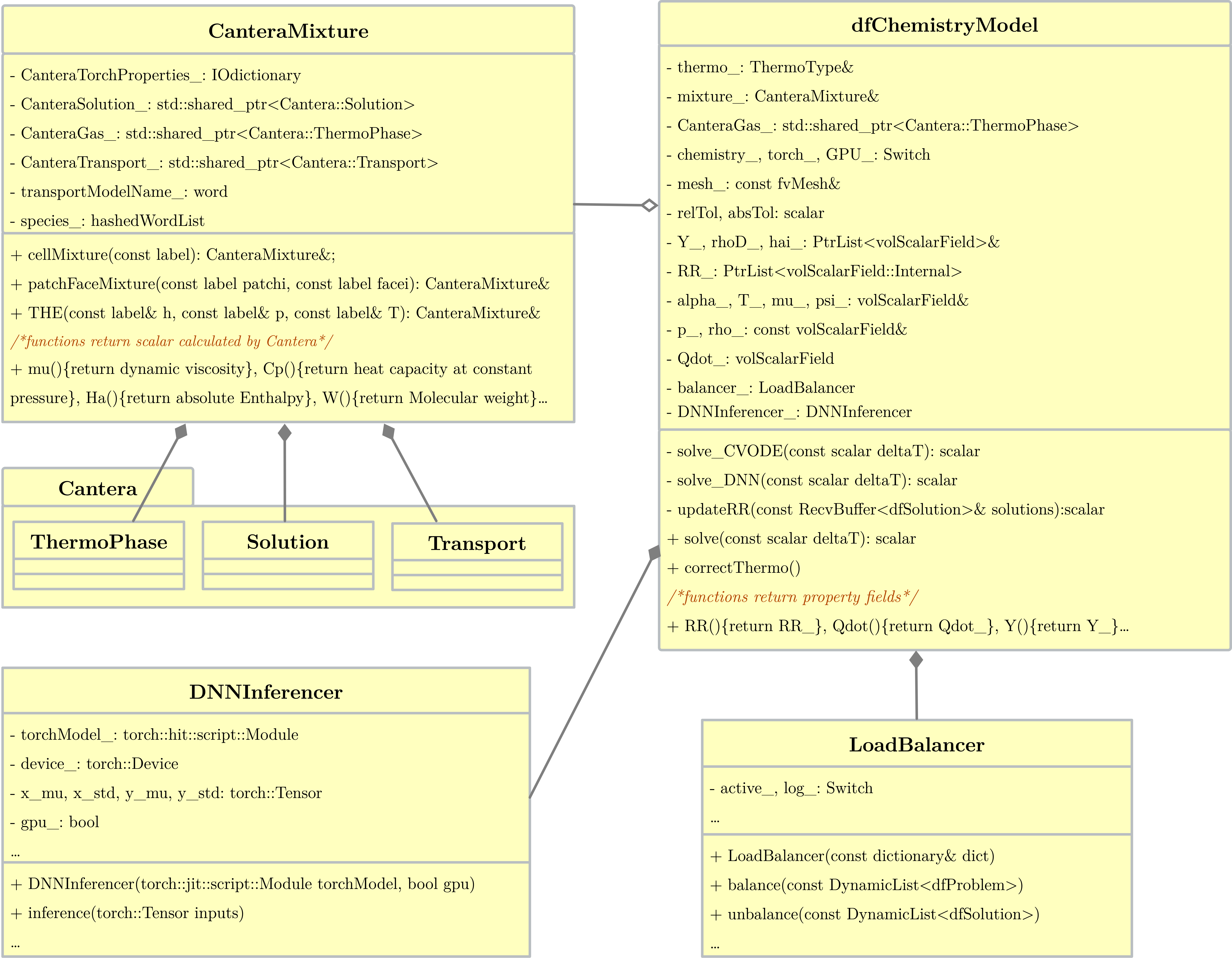}
\caption{The main classes of the thermo-physio-chemistry library in {\em DeepFlame}. The notations ``$-$" and ``$+$" specify the visibility of the class member as public and private.}
\label{fig:classes}
\end{figure}

Figure~\ref{fig:classes} shows the detailed class diagram of the new library in {\em DeepFlame}. The description of four critical classes, {\em CanteraMixture}, {\em dfChemistryModel}, {\em DNNInference} and {\em LoadBalancer} will be described in this section. 

{\em CanteraMixture} is developed to evaluate the thermal properties of Mixture in a single cell with libCantera functions. Note that this class is also adopted in the construction of {\em heThermo} and Chemistry model (see Fig.~{\ref{fig:UML}b), so the operations of which are almost kept identical to the original mixture classes in OpenFOAM. Most of the operations in {\em CanteraMixture} are implemented based on the the attributes {\em CanteraSolution\_}, {\em CanteraGas\_} and {\em CanteraTransport\_}, which are constructed from the original libCantera classes in Fig.~{\ref{fig:classes}}. For example, function {\em mu()} is realised by calling {\em CanteraTransport\_}-$>${\em viscosity()} and function {\em Cp()} is implemented by calling {\em CanteraGas\_}-$>${\em cp\_mass()}. Additionally, to read chemistry mechanisms and user settings , a private attribute {\em CanteraTorchProperties\_} is constructed from the original OpenFOAM I/O controlling class {\em IOdictionary}.

\begin{figure}[!h]
\centering
\includegraphics[width=\textwidth]{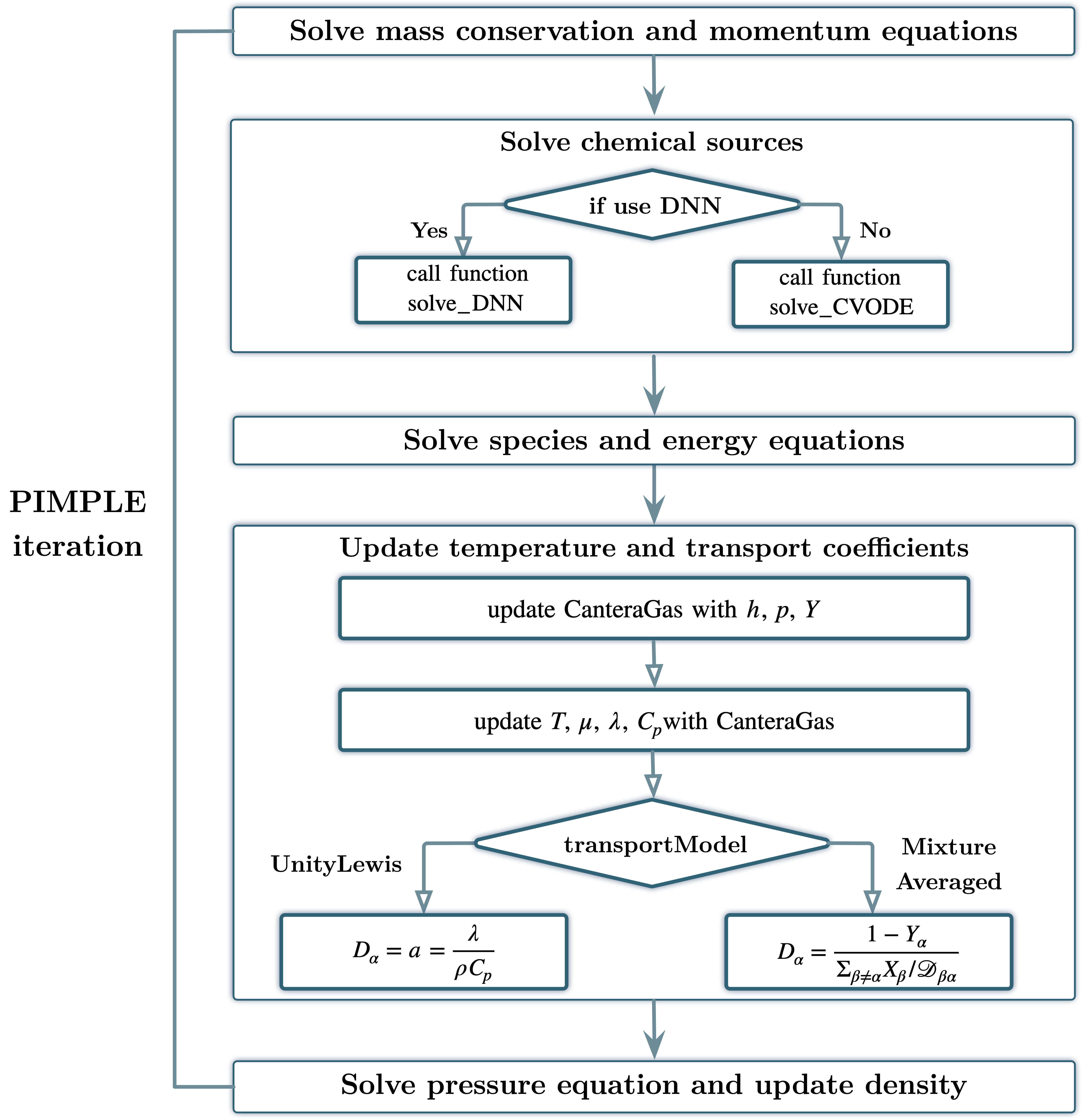}
\caption{A schematic demonstrating the main operations of the {\em dfLowMachFoam}-solver in a single iteration step.}
\label{fig:flowchart}
\end{figure} 

{\em DNNInferencer} is developed to calculate chemical reaction rates via inference of deep neural network. It is constructed with the network, normalisation parameters and a switch for GPU as the DNN inference device. The operation {\em inference} is built to calculate the chemistry source terms with DNN. It mainly consists of three stages: 1) pre-process the input tensor ($p$, $T$, $Y$) by normalisation and Box-Cox transformation; 2) inference the processed tensor with the network; 3) evaluate the mass fractions via the reverse-transformation of stage 1 and finally output the reaction rates. {\em LoadBalancer} is built to balance the chemistry load when using CVODE solver, the detailed implementation is adopted from DLBFoam \cite{RN17}.

Based on the above three classes, the core of {\em DeepFlame}, {\em dfChemistryModel}, can be finally constructed. The thermal and chemistry fields are all contained in this class as the attributes. Also, most of these fields can be received by calling the public operations, such as {\em RR()}, {\em Qdot()}, {\em Y()}, etc. The two public functions {\em solve()} and {\em correctThermo()} are built for solving chemistry source and updating thermal properties. The execution procedure of the two operations is described in Fig.~\ref{fig:flowchart}, which presents the computational algorithm of the {\em dfLowMachFoam}-solver. It can be seen that the conservation equations for mass, momentum, species, energy, and a Poisson equation for pressure are successively solved in an iterative manner. Besides, the other two separate steps, the solving of chemical sources and thermal properties, are achieved by calling the function {\em solve()} and {\em correctThermo()}. The implementation of {\em correctThermo()} can be divided into the following steps: 
construct the attribute {\em CanteraGas\_} with $h$, $p$ and $Y$ obtained from previous solutions; update thermal properties and transport coefficients from the new {\em CanteraGas\_}. Note that the diffusion coefficient ($D_\alpha$) of each species are determined according to the user-defined transport model.

\subsection{Adaptive mesh refinement}\label{subs:AMR}
Adaptive mesh refinement (AMR) is readily available in OpenFOAM as an effective way to reduce computational cost for spatially stiff computations such as involving shock and detonation waves. However, the original AMR algorithm in OpenFOAM named {\em hexRef8} uses the octree refinement and splits each cell in eight child cells (homogeneously halved in three directions). This brings extra computational cost for one- and two- dimensional cases since it will refine cells in invariant directions. Therefore, the original AMR was extended from {\em hexRef8} to {\em hexRef4} in~\cite{amr2015Baniabedalruhman,amr2019load2d} to implement a quadtree refinement for two-dimensional problems. Additionally, multiple refinement criteria was also included for higher flexibility.
%However, it still does not support adaptive refinement for one-dimensional cases. 
In this work, we further extend the AMR capability to one-dimensional via a new mesh cutter named {\em hexRef2}. 
The {\em empty} boundary type (for invariant directions) can be automatically detected, and the corresponding boundary faces will be marked as divisible.
%It can automatically detect the boundary type {\em empty}, and sign those faces as divisible, which is different from {\em hexRef4}. 
During an AMR process, each divisible face is split into two new faces and a new internal face is added to each cell to be refined. 
To keep consistency, we strictly follow the code structure of~\cite{amr2019load2d} and {\em hexRef2} is added as a new derived class to the base class {\em hexRef}. Upon simulation startup, {\em dfDynamicRefineFvMesh} selects the suitable mesh cutter according to the dimension of the case. Additionally, the temporally evolving grid file related I/O is improved to facilitate AMR restart and post-processing. 

\section{Validation and Results}\label{sec:Res}
In this section, the above implementation of {\em DeepFlame} is validated. The solvers described in §~\ref{subs:Solvers} are systematically tested using a broad range of  canonical test cases under different dimension and flow speed conditions. 

\subsection{df0DFoam-solver}\label{subs:0D}

As introduced in §~\ref{subs:GorvEq}, the {\em df0DFoam} is developed to solve zero-dimensional problems, which can be further subdivided into constant-pressure and constant-volume conditions. Figure~\ref{fig:0Dpressure} compares the hydrogen autoignition results obtained by Cantera and {\em df0DFoam} at constant pressure. The chemical mechanism adopted here is developed by Evans et al. \cite{evans1980influence}, containing 8 species and 16 reversible reactions. The neural network is trained according to the method proposed in \cite{zhang2022multi}. In the later hydrogen flame cases, we will also adopt the above mechanism and network if is not specified. In this case, the initial condition (temperature, pressure and equivalence ratio) of the H$_2$/air mixture is set as $T$ = 1400 K, $p$ = 1 atm and $\phi$ = 1. From Fig.~\ref{fig:0Dpressure}, it can be seen that the {\em df0DFoam} can accurately capture the evolution of $T$, $p$ and $Y$, for both the DNN and CVODE integrators. Figure~\ref{fig:0Dvolume} shows constant-volume autoignition results of H$_2$ given by {\em df0DFoam}-solver at $T$ = 1000 K, $p$ = 0.5 atm and $\phi$ = 1. It reveals that the DNN integrator has a satisfying performance even in the condition with a large variation of pressure. To sum up, the cases conducted in this part confirm the validity of the implementation of chemical reaction source terms in {\em DeepFlame}. 

\begin{figure}[!h]
\centering
\includegraphics[width=\textwidth]{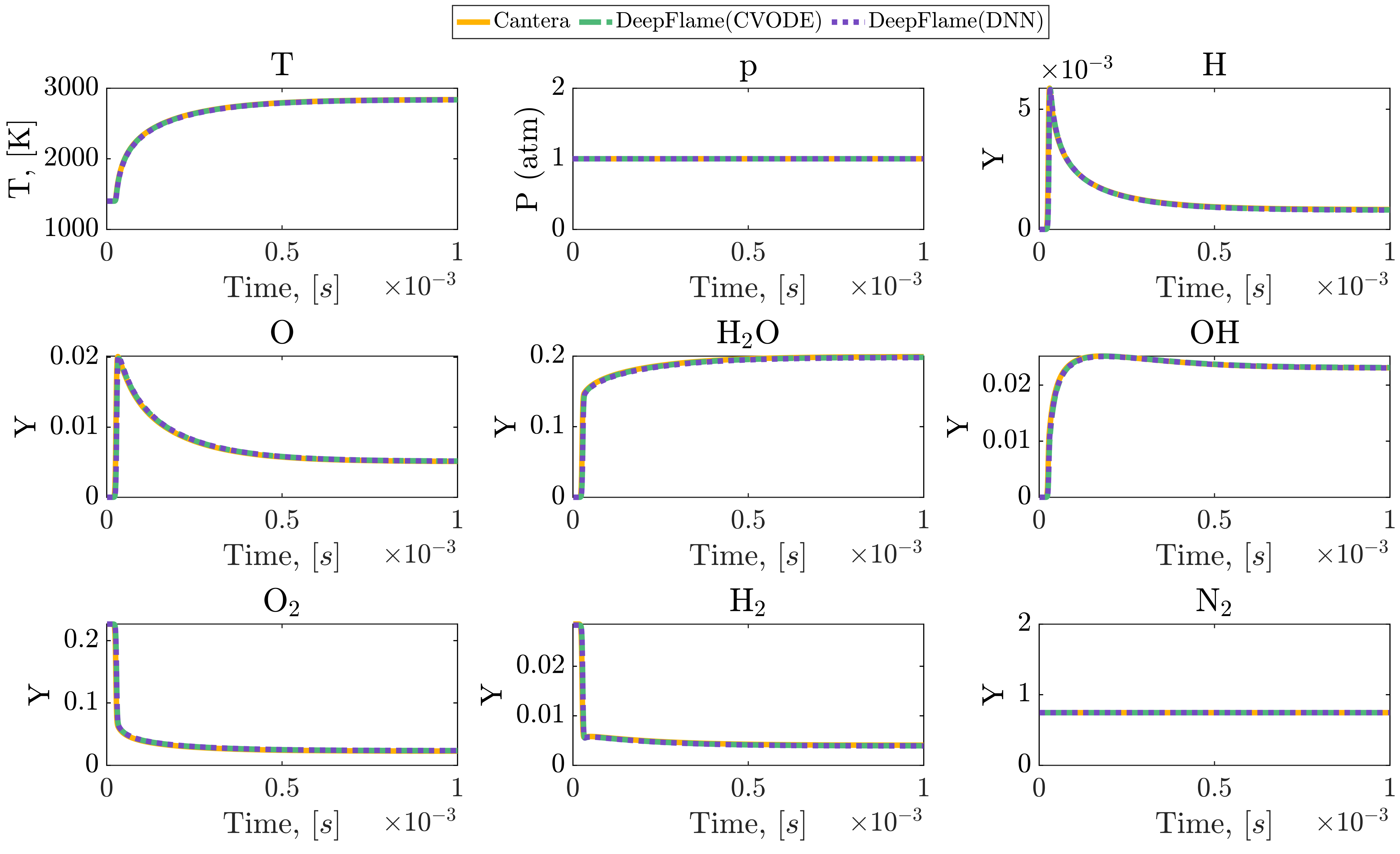}
\caption{Zero-dimensional constant-pressure autoignition results comparison between Cantera and {\em df0DFoam} (with CVODE and DNN integrators).}
\label{fig:0Dpressure}
\end{figure}

\begin{figure}[!h]
\centering
\includegraphics[width=0.6\textwidth]{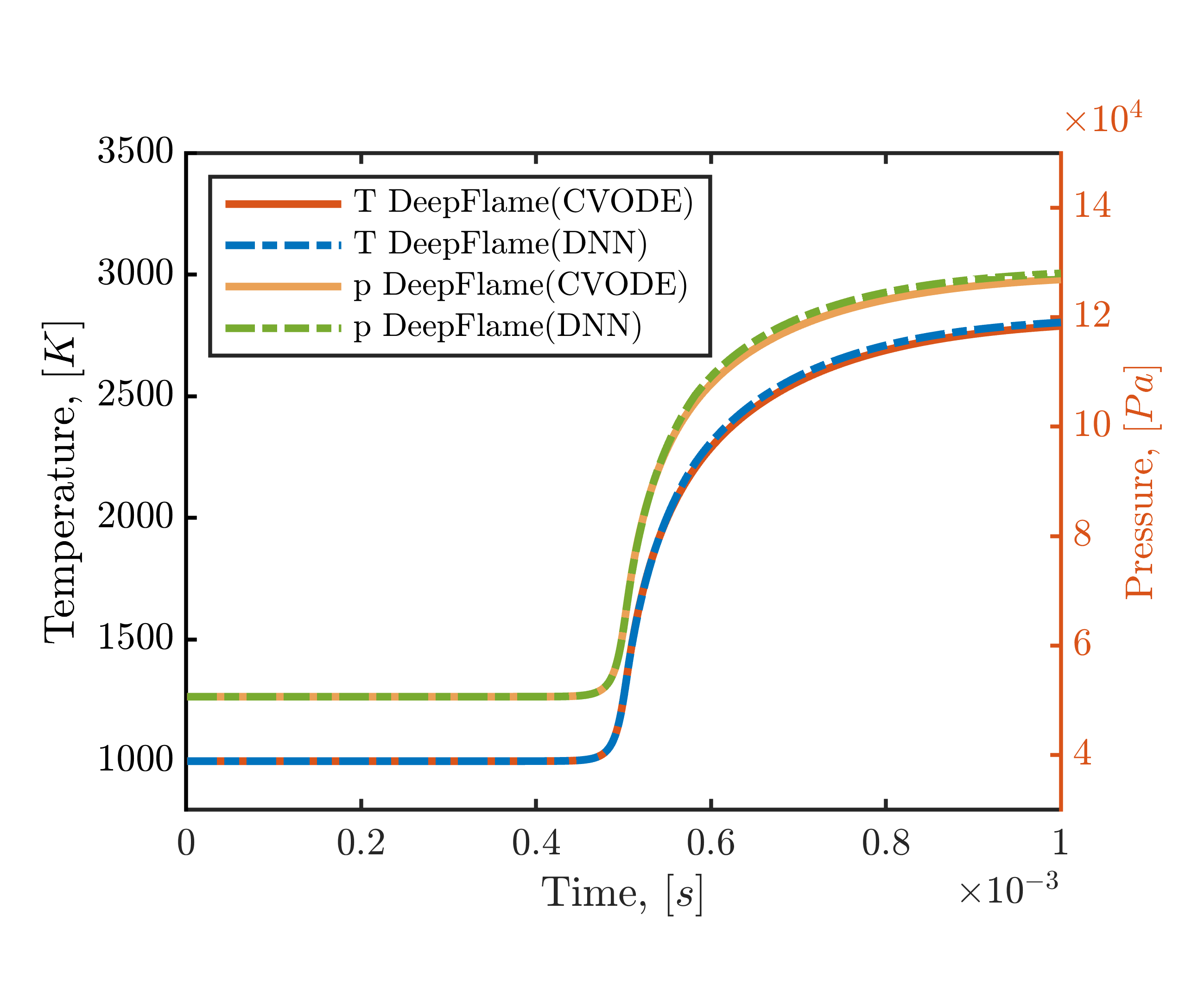}
\caption{Zero-dimensional constant-volume autoignition results given by {\em df0DFoam}.}
\label{fig:0Dvolume}
\end{figure}

Figure~\ref{fig:ignitionDelay} shows the ignition delay times (IDTs, defined using [dT/dt]$_{max}$) versus initial temperature variations for both hydrogen and n-Heptane fuels. For the hydrogen/air mixture at stoichiometric and atmospheric conditions, both the DNN and CVODE chemistry integrators give very close results to the IDTs calculated ated using Cantera for a broad range of temperature conditions. For a more complex fuel n-Heptane (n-C$_7$H$_{16}$) with negative temperature coefficient (NTC) behaviours, {\em df0DFoam} with CVODE also shows very good agreement with the Cantera numerical results and the experimental measurements~\cite{heufer2010determination}. 

\begin{figure}[!h]
\centering
\includegraphics[width=\textwidth]{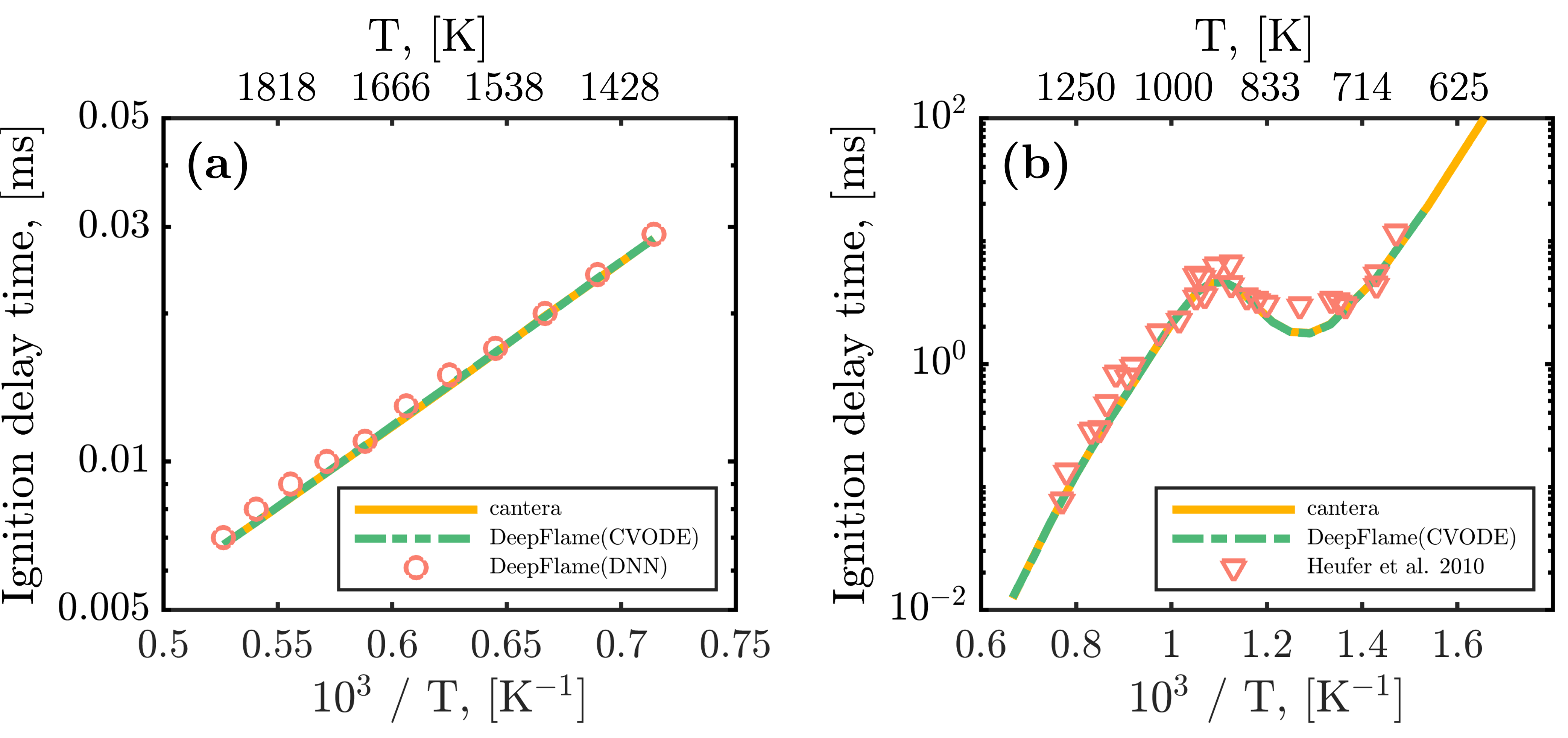}
\caption{Ignition delay time versus temperature for stoichiometric (a) hydrogen/air mixture at atmospheric pressure and (b) n-Heptane/air mixture at 12.5~atm pressure.}
\label{fig:ignitionDelay}
\end{figure}

\subsection{dfLowMachFoam-solver}\label{subs:dfLowMachFoam}

To assess the the implementation of the transport processes, the laminar free-propagating 1D planar premixed flame is first examined. Then, a 2D non-premixed lifted jet flame and a reactive Taylor-Green vortex (TGV) are simulated to validate the capability of the solver for multi-dimensional applications. For all the three cases, hydrogen is adopted as the fuel and the mixture-averaged model is chosen to calculate the diffusion coefficients. Time derivatives are discretised with the implicit Euler scheme. Discretisation of convective and diffusion terms is based on the second-order central differencing scheme. 

\subsubsection{One-Dimensional Planar Flame}

The computational setup of the 1D case is schematically shown in Fig.~\ref{fig:flamefront}a. Except for the Inlet and Outlet, all the side boundaries are defined using the {\em empty} condition (i.e. transverse spatial terms are not calculated), ensuring no fluxes in lateral direction. The domain is initialised using steady-state 1D freely-propagating flame solution from Cantera. The velocity of the inlet mixture is set to be identical to the flame speed given by Cantera so that the flame front can be stabilised inside the domain. The length of the computational domain is according to the solution from Cantera as well, and the cell number is set to ensure the flame front is discretised by at least 30 elements.

\begin{figure}[!h]
\centering
\includegraphics[width=0.7\textwidth]{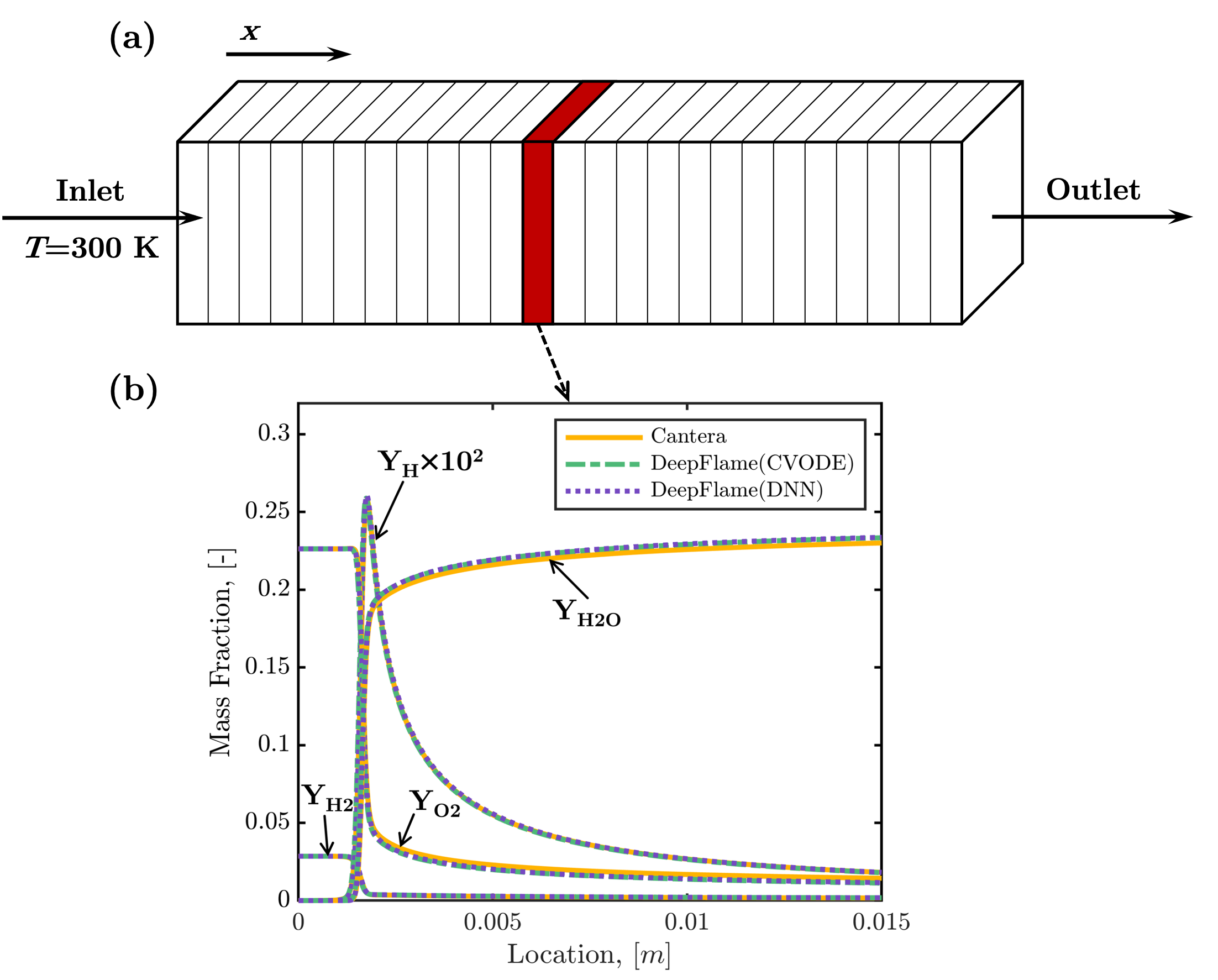}
\caption{(a): Numerical setup of one-dimensional premixed flame. (b): Profiles of mass fractions of main species compared between Cantera and {\em dfLowMachFoam}.}
\label{fig:flamefront}
\end{figure}

\begin{figure}[!h]
\centering
\includegraphics[width=0.8\textwidth]{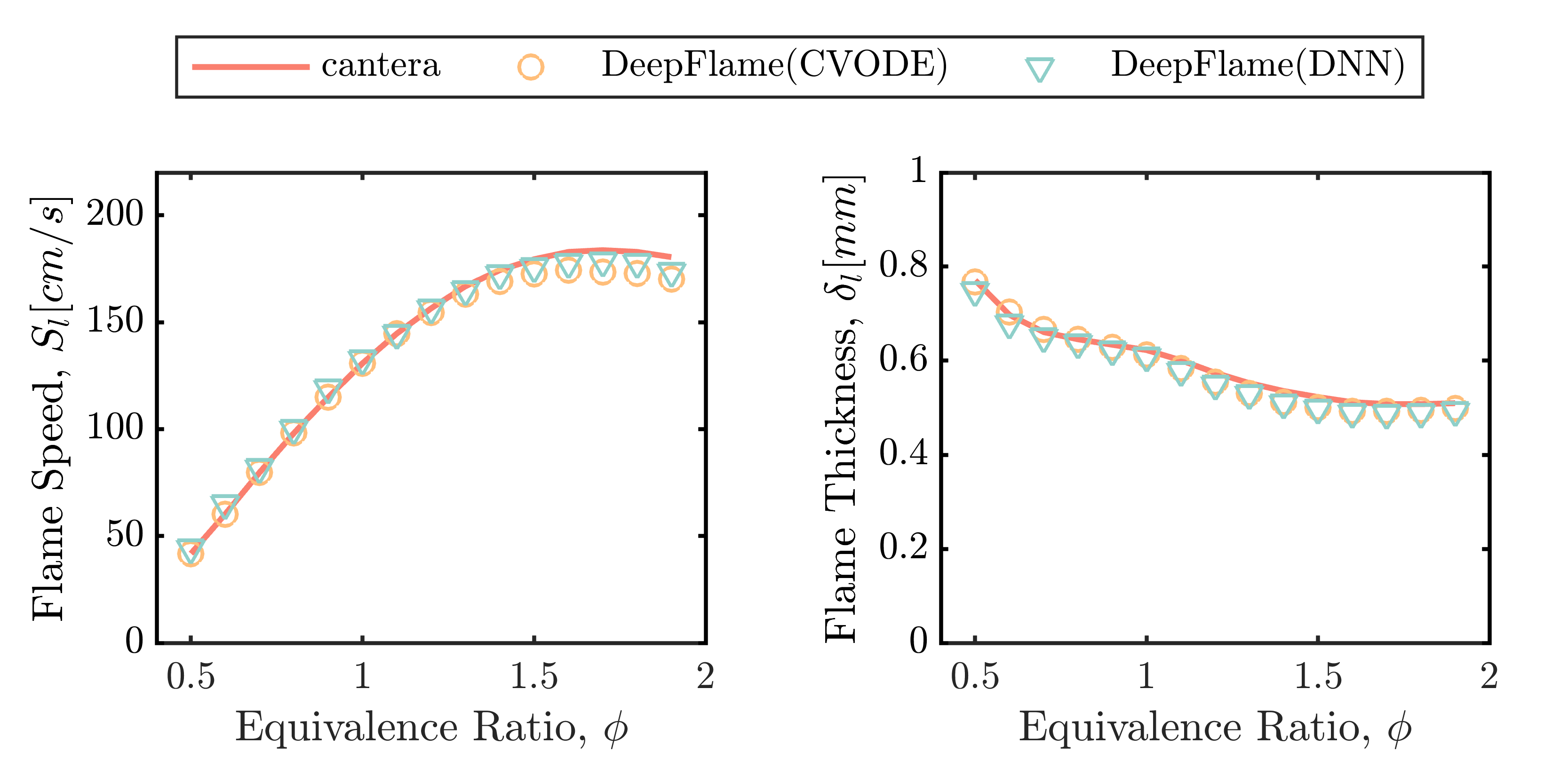}
\caption{Comparison of laminar flame speed (left) and flame thickness (right) of hydrogen/air mixture at different equivalence ratios.}
\label{fig:sl}
\end{figure} 

Figure~\ref{fig:flamefront}b shows that the distributions of the species mass fractions given by {\em dfLowMachFoam}-solver and Cantera exhibit an excellent agreement. This observation can be extended to different equivalence ratios as presented in Fig.~{\ref{fig:sl}. The laminar flame speed ($S_l$) and flame thickness ($\delta_l$) given by {\em dfLowMachFoam} compare well with the Cantera solution for the range of $\phi = $ 0.5 to 1.8. Here $S_l$ is evaluated by subtracting the propagation velocity of flamefront (position of $\nabla T_{max}$) by the inlet velocity and $\delta_l$ is calculated as $(T_b-T_a)/\nabla T_{max}$. The minor deviations in the above comparison can be attributed to the different meshing methods adopted in Cantera and {\em dfLowMachFoam}: the former uses the adaptive grid with high refinement near the flamefront, whereas the latter adopts a static grid. Nevertheless, the cases conducted in this part demonstrate that the convection-diffusion-reaction algorithms implemented in {\em DeepFlame} are stable and accurate.

\subsubsection{Two-Dimensional Jet Flame}

In this subsection, a 2D non-premixed planar jet flame is simulated using {\em dfLowMachFoam} with the DNN and CVODE integrators. As shown in Fig.~\ref{fig:Triple}a, a H$_2$ fuel jet with $T$ =  300 K and $p$ = 1 atm enters from the left end of the rectangle domain, surrounded by an air co-flow. The fuel jet has a diameter of 8~mm and the composition of 25$\%$ H$_2$/75$\%$ N$_2$ in volume. The velocities of the jet and the co-flow are 5 and 1 m/s, respectively. The size of the computational domain is 3 $\times$ 5 cm, discretised by a structured mesh with 300 $\times$ 500 resolution. The initial fields are specified as follows. An ignition region with the temperature of 1400 K is set at the centre of the domain (shaded area in Fig.~\ref{fig:Triple}a); the velocity and mass fractions of the H$_2$ jet decay along the streamwise direction and a 1/7 power law is applied for the velocity profile in the transverse direction. The right figure in  Fig.~\ref{fig:Triple}a depicts the profiles of $u$ and $Y$ along the central line.  

Figure~\ref{fig:Triple}b shows the evolution of the jet flame using heat release rate (HRR) contours. It can be seen that the {\em dfLowMachFoam} gives almost identical results for the DNN and CVODE cases. After ignition, two flame branches develop towards upstream and form the {\em triple\ flame} structures at 4 ms. Here we make a further quantitative comparison between the DNN and CVODE solvers: transverse profiles of species mass fractions in the two triple flames are extracted at the point with maximum heat release rate. As shown in Fig.~\ref{fig:Triple}c, the results obtained by DNN and CVODE show excellent agreement for all major species.

\begin{figure}[!h]
\centering
\includegraphics[width=0.7\textwidth]{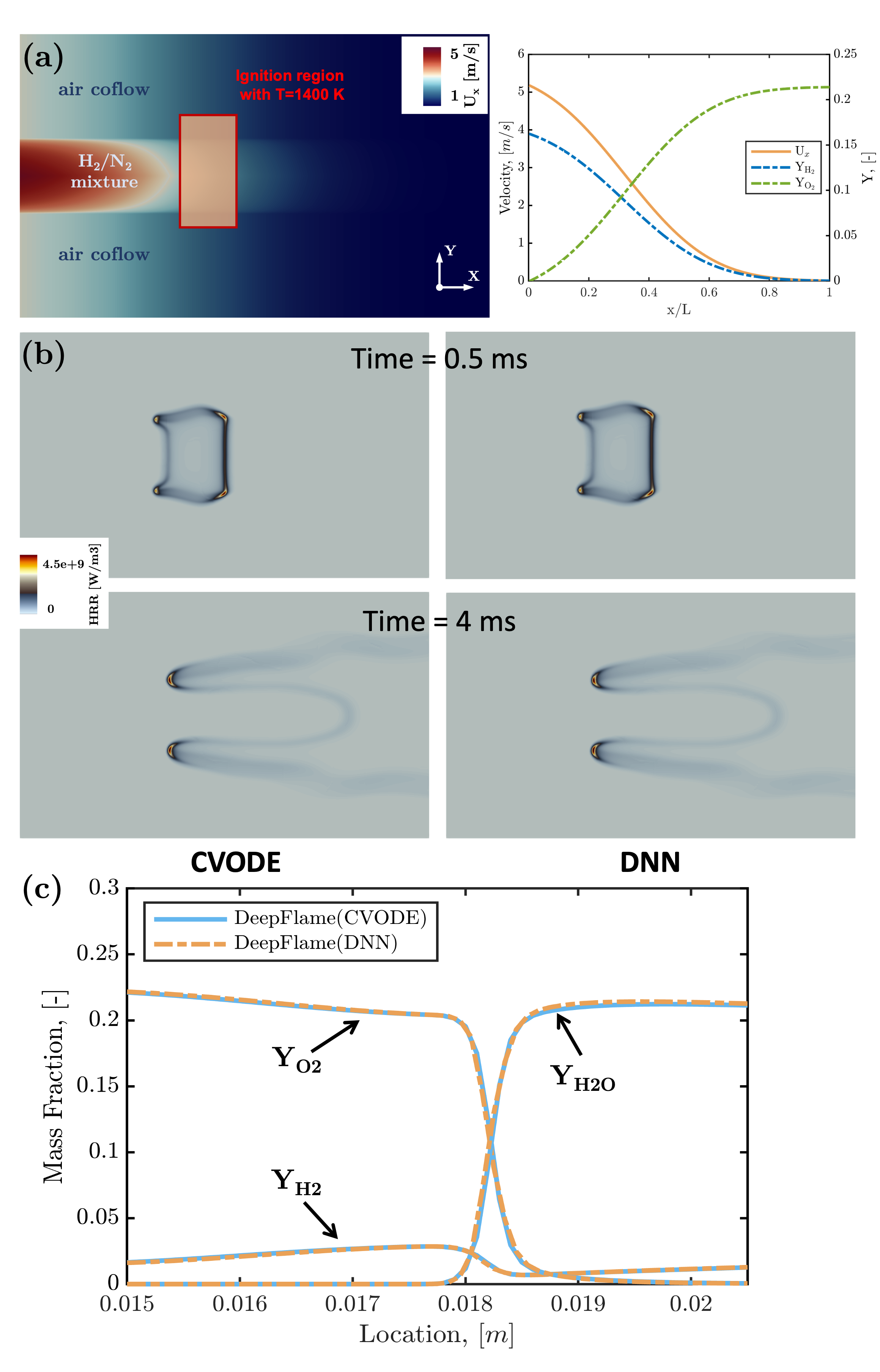}
\caption{Simulation results of the two-dimensional jet flame using {\em dfLowMachFoam}. (a) numerical setup and the initial profiles of velocity and species mass fractions along centreline; (b) Contours of HRR calculated by two ODE integrators at $t$ = 0.5 and 4 ms; (c) Transverse profiles of species mass fractions at the maximum HRR point for the 2D triple flame.}
\label{fig:Triple}
\end{figure}

\subsubsection{Three-Dimensional reactive Taylor-Green Vortex}\label{subs:TGV}
Finally, the performance of {\em dfLowMachFoam} is assessed using a recently established benchmark case for reacting flow DNS codes $-$ 3D Taylor-Green Vortex (TGV) interacting with a non-premixed flame \cite{RN4}. Figure~\ref{fig:TGV1}a shows the cubic computational domain with the edge length of 2$\pi L$ and the initial fields of vorticity magnitude and temperature. All boundaries are set to be  {\em periodic}. The initial condition for the velocity field is given by
\EQ
u_x=u_0\sin(\frac{2\pi x}{L})\cos(\frac{2\pi y}{L})\cos(\frac{2\pi z}{L})\:,
\EN
\EQ
u_y=-u_0\cos(\frac{2\pi x}{L})\sin(\frac{2\pi y}{L})\cos(\frac{2\pi z}{L})\:,
\EN
\EQ
u_z=0\:,
\EN
% \begin{subequations}
% \begin{align}
% u_x&=u_0\sin(\frac{2\pi x}{L})\cos(\frac{2\pi y}{L})\cos(\frac{2\pi z}{L})\\
% u_y&=-u_0\cos(\frac{2\pi x}{L})\sin(\frac{2\pi y}{L})\cos(\frac{2\pi z}{L})\\
% u_z&=0
% \end{align}
% \end{subequations}
where the reference velocity magnitude $u_0$ is set to 1 m/s and the reference length $L$ is set to 1 mm. The chemical mechanism of by Boivin et al. \cite{boivin2011explicit} used in \cite{RN4} involving 9 species and 12 reversible reactions is also adopted here to make a direct comparison. The $x$-direction profiles of the temperature and species are specified following \cite{RN4} as depicted in Fig.~\ref{fig:TGV1}b. The flame is initialised using the equilibrium value of the local thermodynamic state. The more details for the initial setup of this reactive TGV benchmark are given in \cite{RN4}. 

\begin{figure}[!h]
\centering
\includegraphics[width=0.7\textwidth]{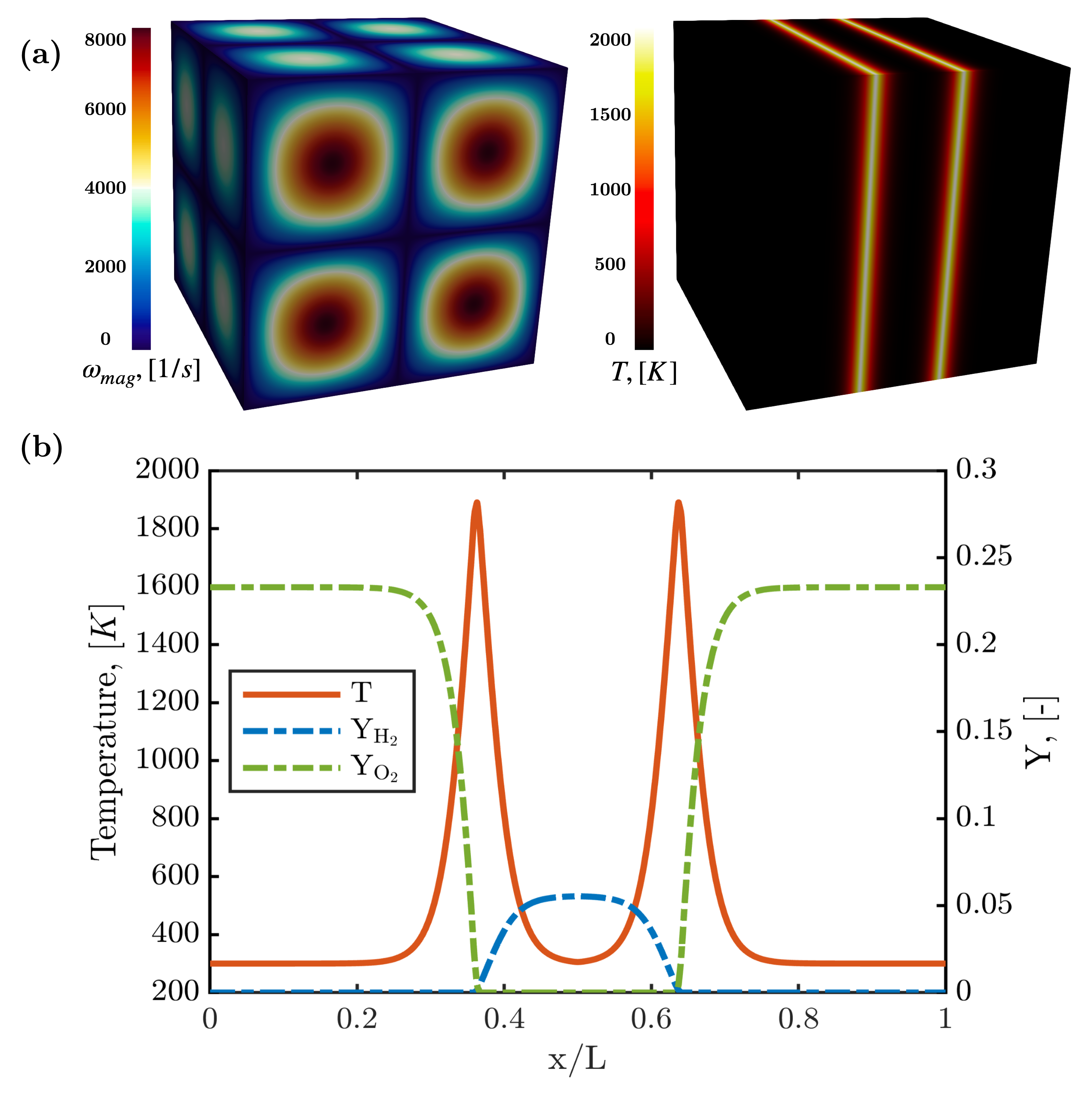}
\caption{(a) Initial contours of vorticity magnitude and temperature for the reactive TGV; (b) Initial profiles of temperature, hydrogen mass fraction and oxygen mass fraction at $y$ = 0.5$L$ and $z$ = 0.5$L$.}
\label{fig:TGV1}
\end{figure}

To fully resolve the flame front, this case is computed on an equidistant grid with 256$^3$ cells. The simulation is carried out for a physical time of $t$ = 0.5 ms (i.e. 2 vortex turnover reference times). Figure~\ref{fig:TGV2}a shows the contours of $Y_{\rm H_2}$ and $T$ calculated using DNN chemistry solver. The accuracy of {\em dfLowMachFoam} in solving complex flow dynamics is demonstrated via the comparison presented in Fig.~\ref{fig:TGV2}b, where the profiles of $Y_{\rm H_2}$ and $T$ at $t$ = 0.5 ms along the $y$-centreline show a quantitatively good agreement between the present and reference (conducted using the code DINO \cite{RN4}) cases.

\begin{figure}[!h]
\centering
\includegraphics[width=0.8\textwidth]{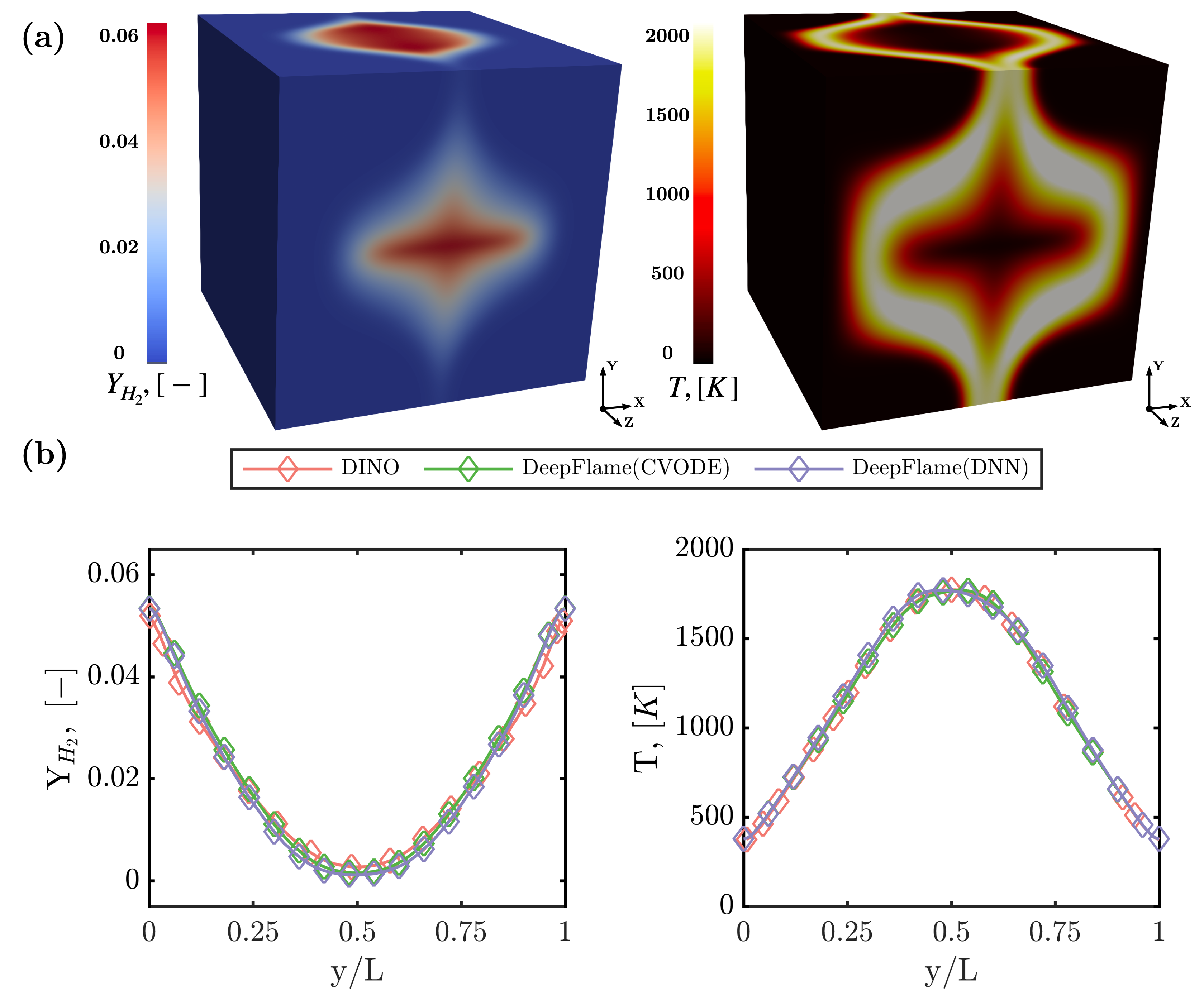}
\caption{(a) Contours of hydrogen mass fractions and temperature calculated with DNN integrators at t = 0.5 ms; (b) Comparison of $Y_{\rm H_2}$ and $T$ at $x$ = 0.5$L$, $z$ = 0.5$L$ and time $t$ = 0.5 ms.}
\label{fig:TGV2}
\end{figure}

\subsection{dfHighSpeedFoam-solver}\label{subs:dfHighSpeedFoam}

In this subsection, we assess the implementation of {\em dfHighSpeedFoam}, in which the original OpenFOAM density-based solver {\em rhoCentralFoam} is extended to consider multi-species transport and reaction also via the {\em dfChemistryModel} class shown earlier in Fig.~\ref{fig:classes}. 
For high-speed supersonic reactive flows, it is necessary to capture the discontinuous structures such as shock and detonation waves, where adaptive meshing becomes essential to ensure both numerical accuracy and efficiency. Thus, the implementation of AMR described in §~\ref{subs:AMR} is also validated here using the following test cases. 

\subsubsection{One-Dimensional Reactive Shock Tube}
First, the multi-component reactive shock tube \cite{shocktube1982Oran} is simulated. The computational domain is a tube of 0.12 m long and filled with the mixture of H$_2$/O$_2$/Ar having a 2/1/7 proportion by mole. The domain is divided into 2400 cells uniformly. The initial condition is set as
\EQ
(T, p, u)=\left\{\begin{array}{lccc}
378.656 \mathrm{~K}, & 7173 \mathrm{~Pa}, & 0 \mathrm{~m} / \mathrm{s}, & x<0.06 \mathrm{~m} \\
748.472 \mathrm{~K}, & 35594 \mathrm{~Pa}, & -487.34 \mathrm{~m} / \mathrm{s}, & x>0.06 \mathrm{~m}
\end{array}\right.
\EN

Supersonic inlet condition is set for the right boundary and solid wall is set for the left boundary. The 9 species and 18 reactions of \cite{mechanism1985H2} mechanism is used in this case. Figure~\ref{fig:reactiveShockTube} shows the distribution of temperature, velocity and mass fraction of H at different times. It can be seen that the results computed using {\rm dfHighSpeedFoam} are in good agreement with the results of Mart\'inez Ferrer et al. \cite{shocktube2014Martinez}. At around $t=190~\mu$s, the reactive flow merges into the shock wave forming a detonative front, which is well captured in this present simulation.

\begin{figure}[!h]
\centering
\includegraphics[width=\textwidth]{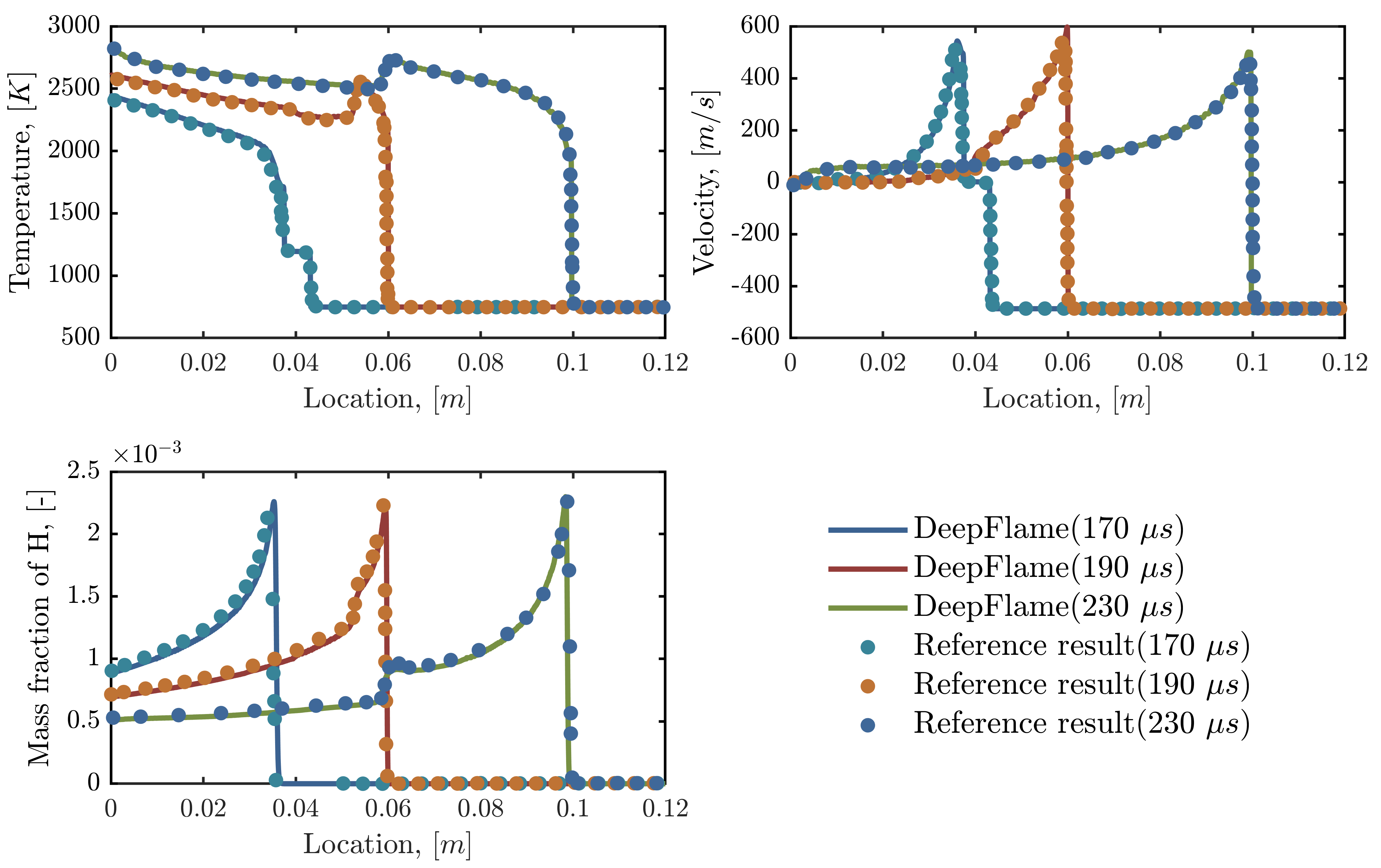}
\caption{Comparisons of simulation results for multi-component reactive shock tube problem between {\em dfHighSpeedFoam} and Ref. \cite{shocktube2014Martinez}: (a) temperature, (b) velocity, (c) mass fraction of H at $t$ = 170 $\mu$s, 190 $\mu$s, and 230 $\mu$s.}
\label{fig:reactiveShockTube}
\end{figure}

\subsubsection{Detonation Propagation Speed}
Detonation propagation can show the coupling of shock wave and chemical reaction.  It contain a complex interaction of the leading shock wave and auto-igniting reaction. This case aims to validate the accuracy of dfHighSpeedFoam in capturing this process and the propagation speed. The length of computational area is 0.5 m. Since adaptive mesh refinement is applied in this case, the coarse cell of 0.8 mm is used in this case. After being refined at the discontinuity, the length of the minimum cell is 0.1 mm since the maxRefinement is set to 3. The domain is filled with homogeneous stoichiometric H$_2$/O$_2$/N$_2$ mixture and the detailed mechanism of 9 species and 21 reactions \cite{mechanism2004Li} is used to calculate combustion. The detonation is ignited by a 2 mm hot spot of 2000 K and 90 atm, which the other area is initialized at 300K and 1atm. 

By recording the position of wave, we can calculate the detonation propagation speed of different equivalent ratio. Figure~\ref{fig:detonation1} show that our result is close to the theoretical result calculated by SDToolbox \cite{SDToolBox}, which can prove the accuracy of this solver.

\begin{figure}[!h]
\centering
\includegraphics[scale=0.5]{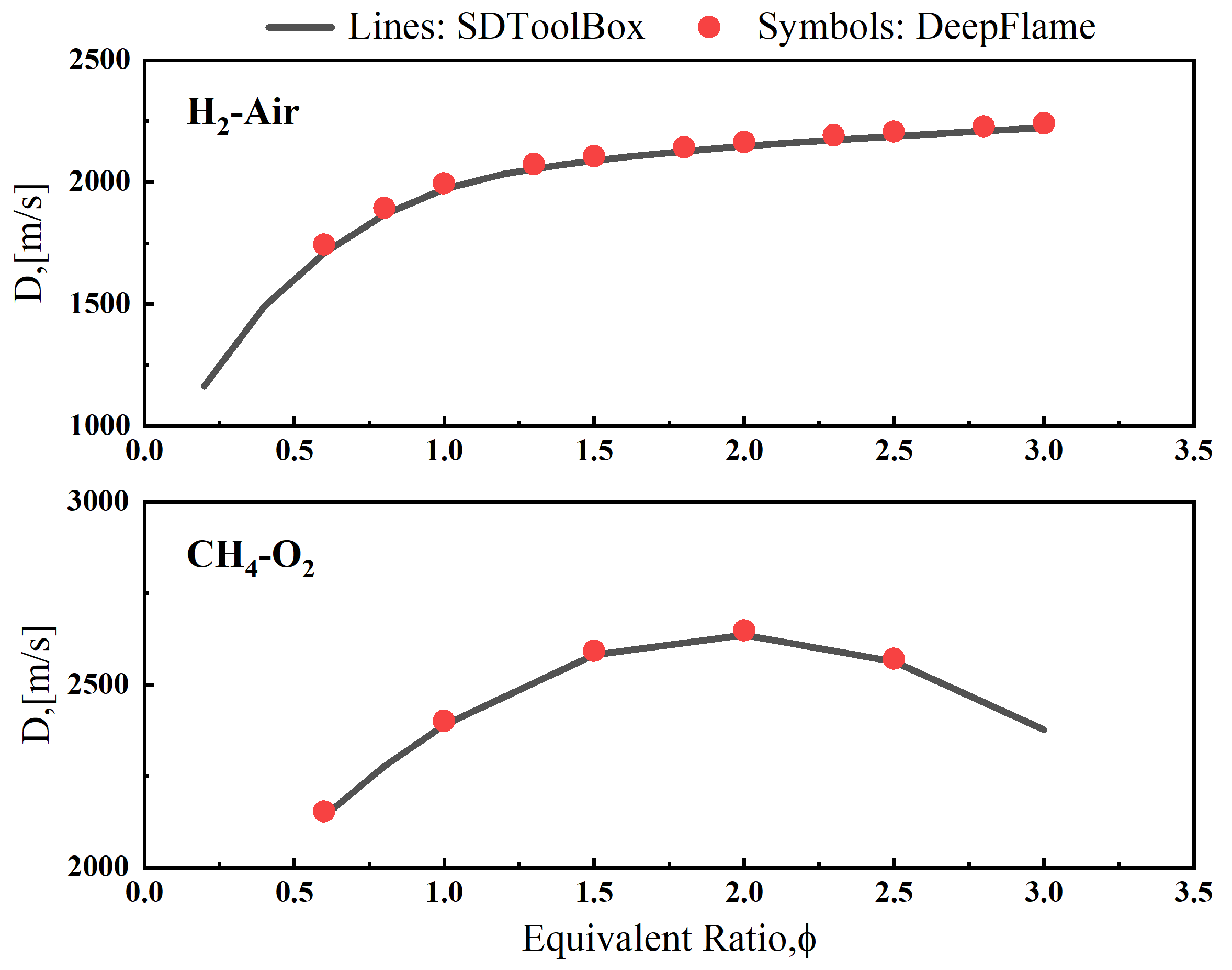}
\caption{Detonation propagation speed versus different equivalent ratio in homogeneous hydrogen/air mixture.}
\label{fig:detonation1}
\end{figure}

\section{Computational Performance}\label{sec:Perf}
As introduced above, machine learning, adaptive mesh refinement and dynamic load balance are adopted in {\em DeepFlame} to speed up reacting flow simulations. In this section, the efficiency improvement given by these approaches will be quantitatively evaluated. First, we will present the speed-up in solving zero-dimensional ignition case when adopt DNN as the chemistry solver. Then the effect of AMR and DLB in accelerating the simulation of one-dimensional detonation is evaluated. Finally, a strong scaling test is conducted for the reactive TGV case to demonstrate the parallel computing efficiency. 

\subsection{Acceleration from Machine Learning}

In this subsection, the performance of the DNN chemistry solver is demonstrated by comparing with the CVODE solver. The computational efficiency of chemistry ODE integrator is evaluated via the execution time in simulating the constant-pressure 0D ignition case (§~\ref{subs:0D}). Figure~\ref{fig:GPUeff} shows the computational time required for different ODE integrators and computing architectures (CPU or GPU). The CPU used here is Intel i7-12700KF and the GPU is NVIDIA RTX3070TI. Note that other than GPU, chips specifically made for AI computations (e.g. Sugon Deep Computing Unit) were also tested but not reported here for conciseness. In general, similar observations to GPU were obtained for the speed-up behaviour – several orders of magnitude acceleration was achieved.

\begin{figure}[!h]
\centering
\includegraphics[width=0.5\textwidth]{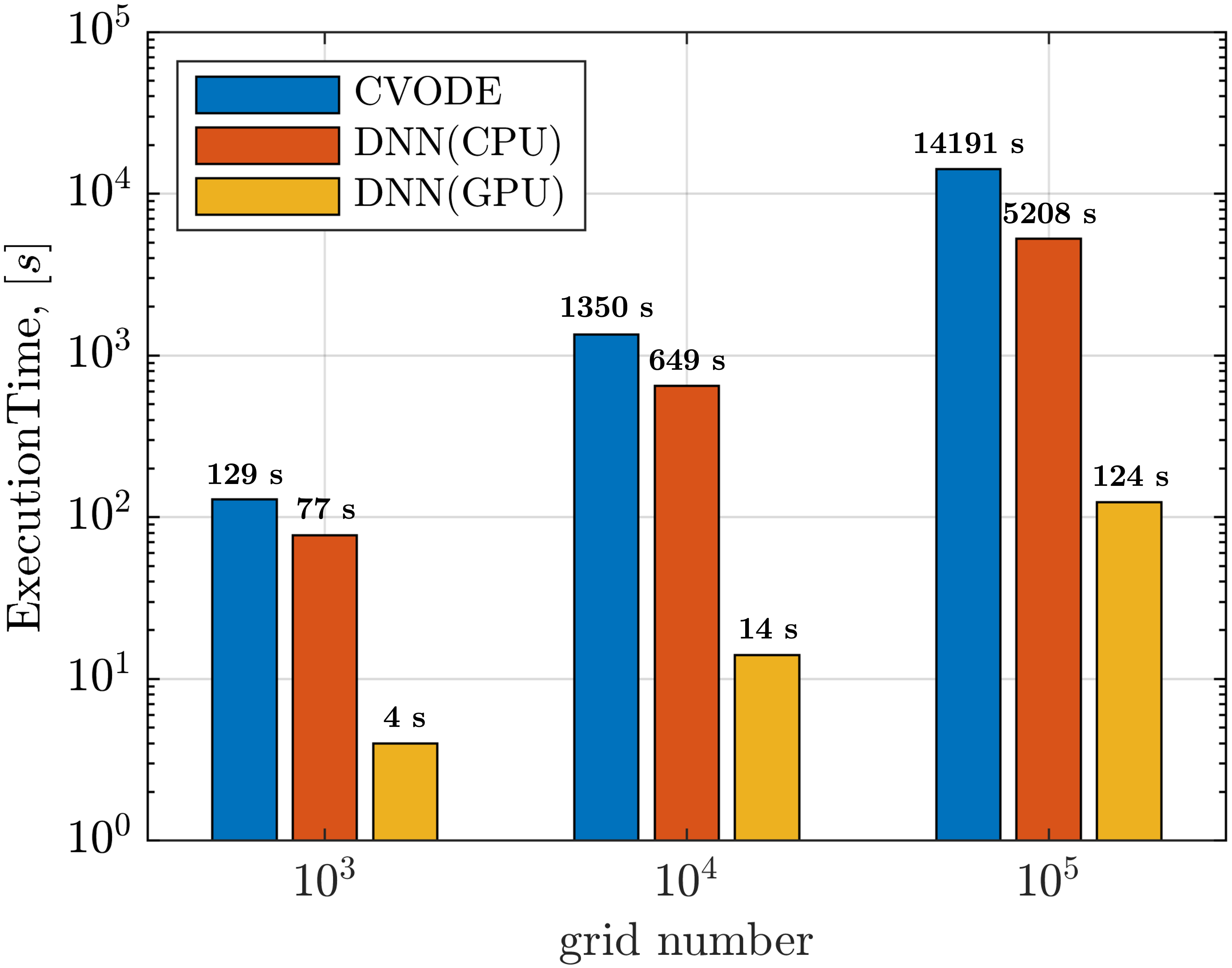}
\caption{Computational time cost in simulating the zero-dimensional ignition problems when adopting different integrators and processing units.}
\label{fig:GPUeff}
\end{figure}

It can be seen in Fig.~\ref{fig:GPUeff} that the computational cost for the given test case is nearly halved on CPU (same as CVODE) and further reduces by two orders of magnitude on GPU when using DNN as the integrator. Additionally, the DNN performance further improves with the increase of the grid number, whereas the CVODE cost linearly scales with the grid number. This speed up given by DNN is essentially attributed to the parallelisation of operators and data, i.e. the solving of chemistry is implemented by matrix addition and multiplication via DNN inference instead of logical iterations. Furthermore, considering the relatively small size of hydrogen/air mechanism, more promising acceleration from DNN can be expected for more complex hydrocarbon fuels.

\subsection{Acceleration from DLB and AMR}

As mentioned earlier, AMR can effectively reduce the simulation cost for high-speed reacting flows. When combined with DLB, the acceleration effect becomes even more significant. As AMR refines the cells in chemically intense areas (near flame, shock or detonation fronts), these spatially clustered cells are likely to be allocated on a small number of processors. This leads to a computational load imbalance, which also varies temporally as the refined region moves within the domain. Therefore, the dynamic balancing technique is almost essential for large-scale multi-dimensional AMR simulations. 

\begin{figure}[!h]
\centering
\includegraphics[scale=1.0]{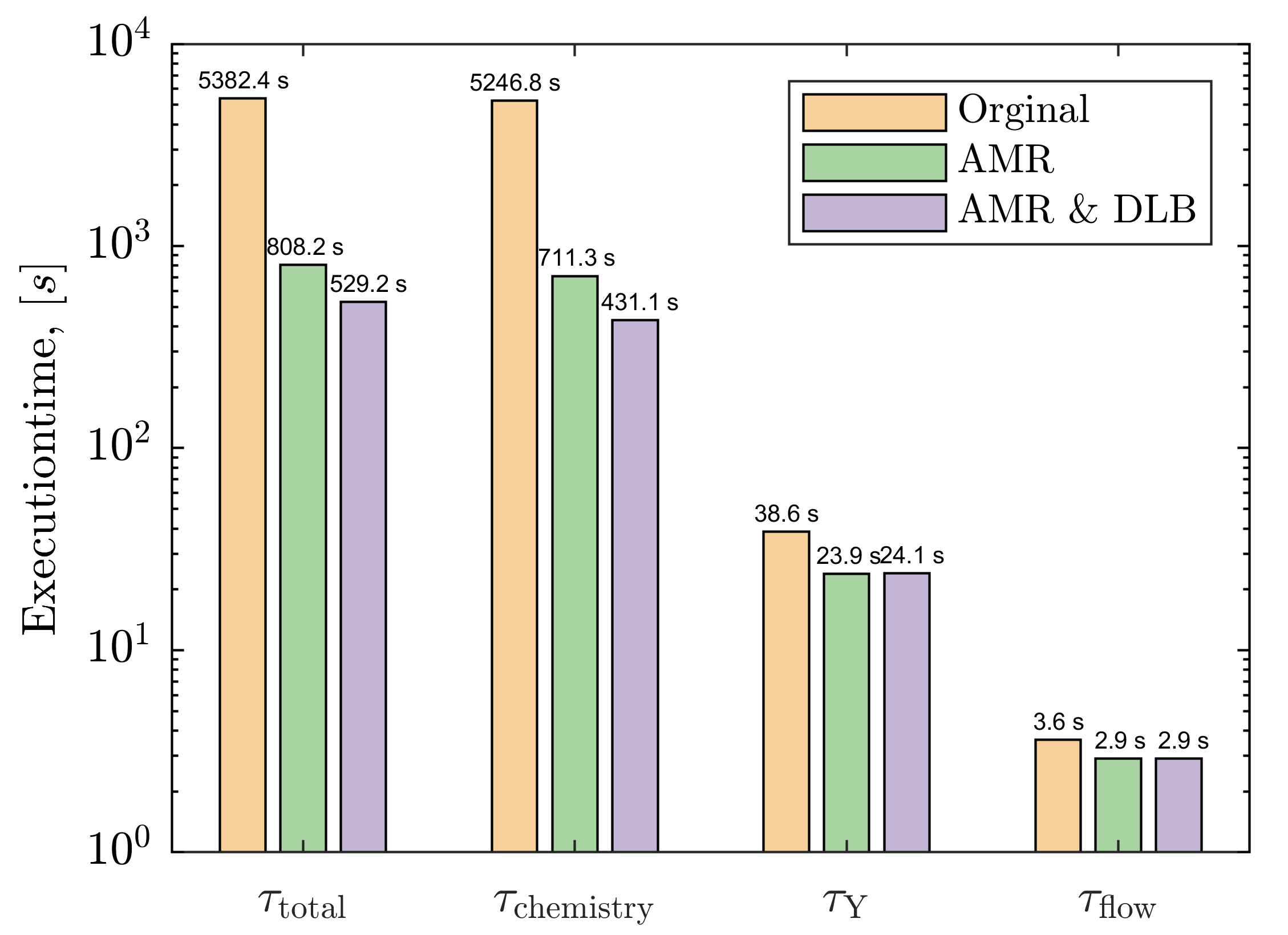}
\caption{Comparison of computational time cost for different calculation parts in simulating one-dimensional detonation problem when closing AMR and DLB, only adapting AMR and adapting both.}
\label{fig:AMR_DLB}
\end{figure}

To demonstrate this effect, Fig.~\ref{fig:AMR_DLB} compares the original, AMR only and AMR+DLB computational times required for a typical 1D detonation simulation using 5 processors. The total time is broken down into 3 parts: chemistry integration, solving species and flow transport equations. It can be observed that chemical source term evaluation accounts for over 95\% of the computation and AMR speeds up the simulation by a factor of about 6.5. When both AMR and DLB are activated, the computation is 10 times faster than the original speed.  

\subsection{Parallel efficiency}

The parallel scalability of {\em DeepFlame} is estimated via a strong scaling test, which is performed by varying the number of processors to solve the same problem. The reactive TGV case introduced in §~\ref{subs:TGV} is adopted here as the benchmark scaling using up to 8192 CPU cores. The machine used in this section is the UK National Supercomputing facility ARCHER2. It is composed of 5860 compute nodes, each with dual AMD EPYCTM 7742 64-core 2.25 GHz processors. 

\begin{table}[h]\footnotesize
\begin{center}
\caption{Comparison of code set-up and performance in simulating reactive TGV. The meanings of the acronyms are given in the text.}
\begin{tabular}{c c c c c c c c c c c c}
\hline
\makecell[c]{Data\\unit} & \makecell[c]{$N$p\\$[-]$}  & \makecell[c]{$N$cores\\$[-]$} & \makecell[c]{$N$it\\$[-]$} & \makecell[c]{$T_{\rm sim}$\\$[\rm{ms_{sim}}]$}& \makecell[c]{$\overline{\Delta t}$\\$\frac{\rm{\mu s_{sim}}}{\rm{iter}}$}& \makecell[c]{TCPU\\$[\rm{h_{CPU}}]$}& \makecell[c]{RCT\\$[\frac{\rm{\mu s_{CPU}}}{\rm{iter\cdot point}}]$}& \makecell[c]{RTTS\\$[\frac{\rm{\mu s_{CPU}}}{\rm{\mu s_{sim}\cdot point}}]$}\\
\hline
YALES2 & 256$^3$ & 384 &  1484 & 2.5 & 1.685 & 923 & 594 & 354  \\
DINO &   256$^3$ & 1024 & 13417 & 2.5 & 0.186 & 2414 & 38 & 660\\
Nek5000  &  256$^3$  & 576 & 3627 & 2.5 & 0.689 & 486 & 201 & 283 \\
DeepFlame  & 256$^3$   & 512 & 2500 & 2.5 & 1 & 1617 & 134 & 134 \\
\hline
\end{tabular}
\label{table:perf}
\end{center}
\end{table}

\begin{figure}[!h]
\centering
\includegraphics[width=0.7\textwidth]{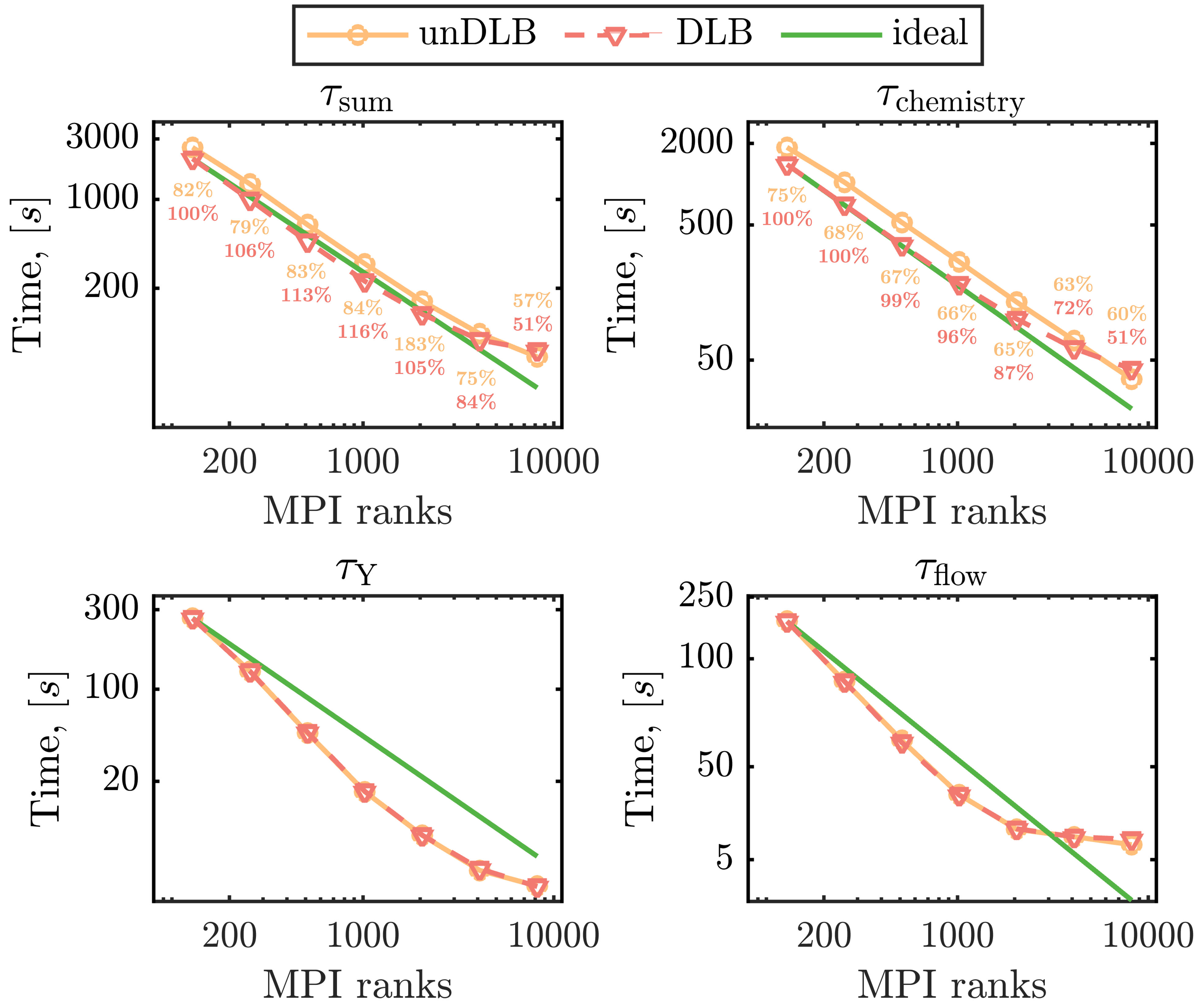}
\caption{Strong scaling for the reactive Taylor-Green vortex setup. Numbers within the top two figures indicate the parallel efficiency.}
\label{fig:TGVscal}
\end{figure}

First, a typical performance comparison between {\em DeepFlame} and the reference codes \cite{RN4} is given in Table~\ref{table:perf}. Here $N$p is the number of grid points, $N$cores is the CPU core number, $N$it is the iteration steps. The subscripts $_{\rm {sim}}$ and $_{\rm {CPU}}$ denote that the variables and units are related to simulation physical time or CPU time. The variable TCPU represents the total CPU time for the simulation and it is computed by multiplying $N$cores and the total CPU time. The Reduced Computational Time (RCT) is calculated with the expression: RCT = TCPU/($N$it $\times$ $N$p). This is introduced to evaluate the CPU time needed to simulate one time-step on a single degree of freedom by using one processor. The CPU time needed to obtain the solution for a single degree of freedom after a specified simulation time $T_{\rm {sim}}$ is represented by Reduced Time to Solution (RTTS), which is calculated as RTTS = TCPU/($T_{\rm{sim}}$ $\times$ $N$p). From the results listed in Table~\ref{table:perf}, {\em DeepFlame} shows the best computational efficiency for the criterion of RTTS. This indicates that for a given duration of flow time, the least CPU time is required for this specific case. The satisfying performance can be mainly attributed to the efficiency of CVODE and DNN chemistry solvers (allowing for a large time-step size) and also the optimised load balancing algorithm. In addition, even for the computation of a single time-step, {\em DeepFlame} gives the second best RCT performance, only slower than the fully structured code DINO with the much faster explicit time-integration. However, it must be noted that all other codes listed in Table~\ref{table:perf} are built with high-order time and space discretisation, whereas {\em DeepFlame} is limited to second order. Nevertheless, this comparison is not intended to claim superiority but to demonstrate the relative computational speed with respect to the state-of-the-art reacting flow codes.

The strong scaling performance is shown in Fig~\ref{fig:TGVscal}. The sum time ($\tau_{sum}$) represents the total elapsed time simulating the reactive TGV from $t$ = 0.5 to 0.6 ms on the given number of CPU cores. The main components of $\tau_{sum}$, $\tau_{chem}$, $\tau_{Y}$ and $\tau_{flow}$, denote the wall-time cost in solving chemistry kinetic equations, species equations and flow equations, respectively. Moreover, both the conditions with and without chemistry dynamic load balance are considered in this section to estimate the performance of the current load balance method. As seen in Fig.~\ref{fig:TGVscal}, $\tau_{chem}$ accounts for more than 70$\%$ of the total time. With the aid of DLB, $\tau_{chem}$ firstly gains a speed-up of about 30$\%$ (for the same MPI rank) but then slows down as the number of CPU cores exceeds about 1000 (i.e. $<16000$ cells per core). This is mainly attributed to the relatively high communication cost as the overhead of DLB at this time. The scaling of $\tau_{Y}$ and $\tau_{flow}$ is super linear up to about 1000 CPU cores. 0However, note that the efficiency in solving flow equations is obviously reduced with the CPU numbers over 5000 ($<3500$ cells per core). This is possibly due to the poor performance of the {\em GAMG}-solver used for solving the Poisson equation of pressure. For this simulation setup, there is an optimum in terms of efficiency at about 16000 cells per core . Further reduction in cell number (i.e. increase CPU number) may benefit the simulation time but at a cost of nonlinear reduction in efficiency.

\section{Conclusion and Future Work}\label{sec:Conclusion}
This study introduced a machine learning empowered CFD platform {\em DeepFlame} for simulating reacting flows. The implementation of this platform is achieved via interfacing  the CFD toolbox OpenFOAM, chemical kinetics software Cantera and deep learning framework Torch. Deferring to all of the previous OpenFOAM-Cantera interface codes, we adopt the simplest nonreactive pure fluid base model and the thermo-physio-chemistry is entirely handled by the {\em DeepFlame} mixture and chemistry objects and functions interfaced with Cantera. Three solvers are developed and thoroughly validated to cover the cases with various flow speeds and physical dimensions. Methods including dynamic load balance (DLB), adaptive mesh refinement (AMR), and deep neural network (DNN) accelerated chemistry solver are available to improve the simulation efficiency.

We also presented a detailed computational performance analysis of {\em DeepFlame}. The validity of the solving algorithms implemented in {\em DeepFlame} is confirmed via a broad range of canonical cases, including 0D ignition, 1D planar flame and reactive shock wave, 2D jet flames, and 3D reactive Taylor-Green vortex. In additional, we also reported the computational efficiency of the code on different computing chips to showcase the acceleration provided by machine learning. Specifically, the DNN chemistry solver shows a 30–50\% speed-up on a CPU due to data and operation vectorisation, and a speed-up of two orders of magnitude on a GPU or DCU due to many-core parallelisation. The simulation of 1D detonation is 10 times faster when adopting AMR and DLB. Finally, the parallel scalability on many CPUs is also reported showing good scaling behaviour compared to the state-of-the-art reactive flow codes  while running on up to tens of thousands of processors. 

Despite the advantages we have introduced in this study, there are still many aspects of {\em DeepFlame} to be improved from the current version: 
\begin{itemize}
    \item The solvers developed in the current platform only focus on the gaseous fuel. Lagrangian/Euler solvers for multi-phase spray reacting flow will be implemented.
    \item The current DNN chemistry solver can only be used on a single GPU/DCU card. In the future we will provide support for the parallel computation of DNN via multi-cards on single and many nodes.
    \item The parallel scalability needs to be further improved to achieve the linear scaling up to more than ten thousands CPU cores. 
\end{itemize}
These features are actively under development on our public repository website and will be released in the future {\em DeepFlame} versions.

%% The Appendices part is started with the command \appendix;
%% appendix sections are then done as normal sections
%% \appendix

\section*{Acknowledgements}
The work of Z.X.C. is supported by the National Science Foundation (Grant No. 52276096) and the Fundamental Research Funds for the Central Universities of China.
Part of the numerical simulations was performed on the High Performance Computing Platform of CAPT of Peking University.

%% \label{}

%% References
%%
%% Following citation commands can be used in the body text:
%% Usage of \cite is as follows:
%%   \cite{key}         ==>>  [#]
%%   \cite[chap. 2]{key} ==>> [#, chap. 2]
%%

%% References with bibTeX database:

\nocite{*}
\bibliographystyle{elsarticle-num}
\bibliography{deepflame}

\begin{thebibliography}{10}
\expandafter\ifx\csname url\endcsname\relax
  \def\url#1{\texttt{#1}}\fi
\expandafter\ifx\csname urlprefix\endcsname\relax\def\urlprefix{URL }\fi
\expandafter\ifx\csname href\endcsname\relax
  \def\href#1#2{#2} \def\path#1{#1}\fi

\bibitem{poinsot2005theoretical}
T.~Poinsot, D.~Veynante, Theoretical and numerical combustion, RT Edwards,
  Inc., 2005.

\bibitem{peters2000turbulent}
N.~Peters, Turbulent combustion, Cambridge university press, 2000.

\bibitem{christo1996artificial}
F.~Christo, A.~Masri, E.~Nebot, Artificial neural network implementation of
  chemistry with pdf simulation of h2/co2 flames, Combustion and Flame 106~(4)
  (1996) 406--427.

\bibitem{blasco1998modelling}
J.~Blasco, N.~Fueyo, C.~Dopazo, J.~Ballester, Modelling the temporal evolution
  of a reduced combustion chemical system with an artificial neural network,
  Combustion and Flame 113~(1-2) (1998) 38--52.

\bibitem{sen2009turbulent}
B.~A. Sen, S.~Menon, Turbulent premixed flame modeling using artificial neural
  networks based chemical kinetics, Proceedings of the Combustion Institute
  32~(1) (2009) 1605--1611.

\bibitem{wan2020chemistry}
K.~Wan, C.~Barnaud, L.~Vervisch, P.~Domingo, Chemistry reduction using machine
  learning trained from non-premixed micro-mixing modeling: Application to dns
  of a syngas turbulent oxy-flame with side-wall effects, Combustion and Flame
  220 (2020) 119--129.

\bibitem{yao2022gradient}
S.~Yao, A.~Kronenburg, A.~Shamooni, O.~Stein, W.~Zhang, Gradient boosted
  decision trees for combustion chemistry integration, Applications in Energy
  and Combustion Science 11 (2022) 100077.

\bibitem{zhang2022multi}
T.~Zhang, Y.~Yi, Y.~Xu, Z.~X. Chen, Y.~Zhang, E.~Weinan, Z.-Q.~J. Xu, A
  multi-scale sampling method for accurate and robust deep neural network to
  predict combustion chemical kinetics, Combustion and Flame 245 (2022) 112319.

\bibitem{chen2021application}
Z.~X. Chen, S.~Iavarone, G.~Ghiasi, V.~Kannan, G.~D’Alessio, A.~Parente,
  N.~Swaminathan, Application of machine learning for filtered density function
  closure in mild combustion, Combustion and Flame 225 (2021) 160--179.

\bibitem{chi2022efficient}
C.~Chi, X.~Xu, D.~Th{\'e}venin, Efficient premixed turbulent combustion
  simulations using flamelet manifold neural networks: A priori and a
  posteriori assessment, Combustion and Flame 245 (2022) 112325.

\bibitem{perry2022co}
B.~A. Perry, M.~T.~H. de~Frahan, S.~Yellapantula, Co-optimized machine-learned
  manifold models for large eddy simulation of turbulent combustion, Combustion
  and Flame 244 (2022) 112286.

\bibitem{zhang2020large}
Y.~Zhang, S.~Xu, S.~Zhong, X.-S. Bai, H.~Wang, M.~Yao, Large eddy simulation of
  spray combustion using flamelet generated manifolds combined with artificial
  neural networks, Energy and AI 2 (2020) 100021.

\bibitem{abadi2016tensorflow}
M.~Abadi, P.~Barham, J.~Chen, Z.~Chen, A.~Davis, J.~Dean, M.~Devin,
  S.~Ghemawat, G.~Irving, M.~Isard, et~al., $\{$TensorFlow$\}$: a system for
  $\{$Large-Scale$\}$ machine learning, in: 12th USENIX symposium on operating
  systems design and implementation (OSDI 16), 2016, pp. 265--283.

\bibitem{collobert2011torch7}
R.~Collobert, K.~Kavukcuoglu, C.~Farabet, Torch7: A matlab-like environment for
  machine learning, in: BigLearn, NIPS workshop, no. CONF, 2011.

\bibitem{opencfd2009open}
O.~OpenCFD, The open source cfd toolbox, User Guide, OpenCFD Ltd 770.

\bibitem{goodwin2002cantera}
D.~G. Goodwin, Cantera c++ user’s guide, California Institute of Technology
  (2002) 32.

\bibitem{curtiss1949transport}
C.~F. Curtiss, J.~O. Hirschfelder, Transport properties of multicomponent gas
  mixtures, The Journal of Chemical Physics 17~(6) (1949) 550--555.

\bibitem{kee2005chemically}
R.~J. Kee, M.~E. Coltrin, P.~Glarborg, Chemically reacting flow: theory and
  practice, John Wiley \& Sons, 2005.

\bibitem{ern1994lecture}
A.~Ern, V.~Giovangigli, Lecture notes in physics, Multicomponent transport
  algorithms 24.

\bibitem{strang1968construction}
G.~Strang, On the construction and comparison of difference schemes, SIAM
  journal on numerical analysis 5~(3) (1968) 506--517.

\bibitem{lu2009toward}
T.~Lu, C.~K. Law, Toward accommodating realistic fuel chemistry in large-scale
  computations, Progress in Energy and Combustion Science 35~(2) (2009)
  192--215.

\bibitem{RN12}
T.~Li, J.~Pan, F.~Kong, B.~Xu, X.~Wang, A quasi-direct numerical simulation
  solver for compressible reacting flows, Computers \& Fluids 213.
\newblock \href {http://dx.doi.org/10.1016/j.compfluid.2020.104718}
  {\path{doi:10.1016/j.compfluid.2020.104718}}.

\bibitem{RN5}
Q.~Yang, P.~Zhao, H.~Ge, reactingfoam-sci: An open source cfd platform for
  reacting flow simulation, Computers \& Fluids 190 (2019) 114--127.
\newblock \href {http://dx.doi.org/10.1016/j.compfluid.2019.06.008}
  {\path{doi:10.1016/j.compfluid.2019.06.008}}.

\bibitem{RN13}
F.~Zhang, H.~Bonart, T.~Zirwes, P.~Habisreuther, H.~Bockhorn, N.~Zarzalis,
  Direct Numerical Simulation of Chemically Reacting Flows with the Public
  Domain Code OpenFOAM, 2015, book section Chapter 16, pp. 221--236.
\newblock \href {http://dx.doi.org/10.1007/978-3-319-10810-0_16}
  {\path{doi:10.1007/978-3-319-10810-0_16}}.

\bibitem{RN14}
D.~Zhou, H.~Zhang, S.~Yang, A robust reacting flow solver with computational
  diagnostics based on openfoam and cantera, Aerospace 9~(2).
\newblock \href {http://dx.doi.org/10.3390/aerospace9020102}
  {\path{doi:10.3390/aerospace9020102}}.

\bibitem{hindmarsh2005sundials}
A.~C. Hindmarsh, P.~N. Brown, K.~E. Grant, S.~L. Lee, R.~Serban, D.~E.
  Shumaker, C.~S. Woodward, Sundials: Suite of nonlinear and
  differential/algebraic equation solvers, ACM Transactions on Mathematical
  Software (TOMS) 31~(3) (2005) 363--396.

\bibitem{box1964analysis}
G.~E. Box, D.~R. Cox, An analysis of transformations, Journal of the Royal
  Statistical Society: Series B (Methodological) 26~(2) (1964) 211--243.

\bibitem{RN17}
B.~Tekgül, P.~Peltonen, H.~Kahila, O.~Kaario, V.~Vuorinen, Dlbfoam: An
  open-source dynamic load balancing model for fast reacting flow simulations
  in openfoam, Computer Physics Communications 267.
\newblock \href {http://dx.doi.org/10.1016/j.cpc.2021.108073}
  {\path{doi:10.1016/j.cpc.2021.108073}}.

\bibitem{amr2015Baniabedalruhman}
A.~Baniabedalruhman, Dynamic meshing around fluid-fluid interfaces with
  applications to droplet tracking in contraction geometries, Dissertation,
  Michigan Technological University.

\bibitem{amr2019load2d}
D.~Rettenmaier, D.~Deising, Y.~Ouedraogo, E.~Gjonaj, H.~D. Gersem, D.~Bothe,
  C.~Tropea, H.~Marschall, Load balanced 2d and 3d adaptive mesh refinement in
  openfoam, SoftwareX 10 (2019) 100317.
\newblock \href {http://dx.doi.org/10.1016/j.softx.2019.100317}
  {\path{doi:10.1016/j.softx.2019.100317}}.

\bibitem{evans1980influence}
J.~S. Evans, C.~J. Schexnayder~Jr, Influence of chemical kinetics and
  unmixedness on burning in supersonic hydrogen flames, AIAA journal 18~(2)
  (1980) 188--193.

\bibitem{heufer2010determination}
K.~Heufer, H.~Olivier, Determination of ignition delay times of different
  hydrocarbons in a new high pressure shock tube, Shock Waves 20~(4) (2010)
  307--316.

\bibitem{RN4}
A.~Abdelsamie, G.~Lartigue, C.~E. Frouzakis, D.~Thévenin, The taylor–green
  vortex as a benchmark for high-fidelity combustion simulations using low-mach
  solvers, Computers \& Fluids 223.
\newblock \href {http://dx.doi.org/10.1016/j.compfluid.2021.104935}
  {\path{doi:10.1016/j.compfluid.2021.104935}}.

\bibitem{boivin2011explicit}
P.~Boivin, C.~Jim{\'e}nez, A.~L. S{\'a}nchez, F.~A. Williams, An explicit
  reduced mechanism for h2--air combustion, Proceedings of the Combustion
  Institute 33~(1) (2011) 517--523.

\bibitem{shocktube1982Oran}
E.~S. Oran, T.~R. Young, J.~P. Boris, A.~Cohen, Weak and strong ignition. i.
  numerical simulations of shock tube experiments, Combustion and Flame 48
  (1982) 135--148.

\bibitem{mechanism1985H2}
R.~J. Kee, J.~F. Grcar, M.~D. Smooke, J.~A. Miller, E.~Meeks, Premix: A fortran
  program for modeling steady laminar one-dimensional premixed flames, Sandia
  National Laboratories.

\bibitem{shocktube2014Martinez}
P.~J. Martínez~Ferrer, R.~Buttay, G.~Lehnasch, A.~Mura, A detailed
  verification procedure for compressible reactive multicomponent
  navier–stokes solvers, Computers \& Fluids 89 (2014) 88--110.

\bibitem{mechanism2004Li}
J.~Li, Z.~Zhao, A.~Kazakov, F.~L. Dryer, An updated comprehensive kinetic model
  of hydrogen combustion, International Journal of Chemical Kinetics 36 (2004)
  566--575.

\bibitem{SDToolBox}
J.~E. Shepherd, Shock $\&$ detonation toolbox,
  \url{https://shepherd.caltech.edu/EDL/publicresources}.

\end{thebibliography}

%% Authors are advised to submit their bibtex database files. They are
%% requested to list a bibtex style file in the manuscript if they do
%% not want to use elsarticle-num.bst.

%% References without bibTeX database:

% \begin{thebibliography}{00}

%% \bibitem must have the following form:
%%   \bibitem{key}...
%%

% \bibitem{}

% \end{thebibliography}

\end{document}